\documentclass[useAMS,usenatbib]{mn2e}
\usepackage{graphicx}
\usepackage{pdflscape}
\usepackage{rotating}
\usepackage{amssymb}

\usepackage{mathrsfs}

\usepackage{amsxtra}
\usepackage{textcomp}
\usepackage{afterpage}
\usepackage{cases}
\usepackage{color}
\usepackage{multirow}
\usepackage{natbib}
\usepackage{times}
\usepackage{natbib}
\usepackage{amsmath}
\usepackage{amssymb}
\usepackage{textcomp}
\usepackage[normalem]{ulem}
\usepackage{multirow}
\usepackage{capt-of}
\usepackage{array}
\usepackage{tikz}
\usetikzlibrary{trees}
\usepackage[colorinlistoftodos]{todonotes}
\usepackage{cases}
\usepackage[export]{adjustbox}

\def \aap  {A\&A}

\def \aj  {AJ}
\def \apj  {ApJ}
\def \apjs  {ApJS}

\def \mnras {MNRAS}
\def \nat {Nature}
\def \apjl {ApJL}

\title[Projected phase--space of EDisCS]{The time delay between star formation quenching and morphological transformation of galaxies in clusters: a phase--space view of EDisCS}
\author[Kshitija Kelkar et al.]{Kshitija Kelkar$^{1}$\thanks{E-mail:
kshitija@rri.res.in (KK) }, Meghan E. Gray$^{2}$, Alfonso Arag\'{o}n-Salamanca$^{2}$, \newauthor Gregory Rudnick$^{3}$, Yara L. Jaff\'{e}$^{4}$, Pascale Jablonka$^{5,6}$, John Moustakas$^{7}$, \newauthor Bo Milvang-Jensen$^{8,9}$ \\
$^{1}$Raman Research Institute, Bangalore 560080, India\\
$^{2}$School of Physics \& Astronomy, University of Nottingham, Nottingham NG7 2RD, UK\\
$^{3}$Department of Physics and Astronomy, University of Kansas, KS 66045-7582 \\
$^{4}$Instituto de F\'{i}sica y Astronom\'{i}a, Facultad de Ciencias, Universidad de Valpara\'{i}so, Gran Breta\~na 1111, 5030 Casilla, Valpara\'{i}so, Chile \\
$^{5}$Laboratoire d'Astrophysique, \'{E}cole Polytechnique F\'{e}d\'{e}rale de Lausanne (EPFL), Observatoire de Sauverny, 1290, Versoix, Switzerland\\
$^{6}$ GEPI, Observatoire de Paris, CNRS UMR 8111, Université Paris Diderot, 92125, Meudon Cedex, France\\
$^{7}$Department of Physics \& Astronomy, Siena College, NY 12211\\
$^{8}$Dark Cosmology Centre, Niels Bohr Institute, University of Copenhagen, Juliane Maries Vej 30, 2100, Copenhagen, Denmark\\
$^{9}$Cosmic Dawn Center (DAWN), Niels Bohr Institute, University of Copenhagen, Juliane Maries Vej 30, 2100 Copenhagen, Denmark}

\begin{document}

\date{Accepted xxx. Received xxx; in original form xxx}

\pagerange{\pageref{firstpage}--\pageref{lastpage}} \pubyear{xxxx}

\maketitle

\label{firstpage}

\begin{abstract}

We explore the possible effect of cluster environments on the structure and star formation histories of galaxies by analysing the projected phase--space (PPS) of intermediate-redshift clusters ($0.4\leqslant z \leqslant0.8$). {\em HST} $I-$band imaging data from the ESO Distant Cluster Survey (EDisCS) allow us to measure deviations of the galaxies' light distributions from symmetric and smooth profiles using two parameters, $A_{\rm res}$ (`asymmetry') and $RFF$ (residual flux fraction or `roughness').  Combining these structural parameters with age-sensitive spectral indicators ($H_{\delta\rm{A}}$, $H_{\gamma\rm{A}}$ and $D_{n}4000$), we establish that in all environments younger star-forming galaxies of all morphologies are `rougher' and more asymmetric than older, more quiescent ones. Combining a subset of the EDisCS clusters we construct a stacked PPS diagram and find a significant correlation between the position of the galaxies on the PPS and their stellar ages, irrespective of their morphology. We also observe an increasing
fraction of galaxies with older stellar populations towards the cluster core, while the galaxies’ structural parameters  ($A_{\rm res}$ and $RFF$) do not seem to segregate strongly with PPS. These results may imply that, under the possible influence of their immediate cluster environment, galaxies have their star formation suppressed earlier, while their structural transformation happens on a longer timescale as they accumulate and age in the cluster cores.

\end{abstract}


\section{Introduction}
\label{secintro}

Observations of galaxy clusters since $z\sim2$ have established that the cluster environment is anti-correlated with star formation in galaxies \citep{kauffmann04,baldry06,balogh07,haines13,kovac14}. This relation, when taken into account with the discernible colour bimodality in clusters, advocates the plausible role environment might be playing in transforming the blue star-forming galaxies to the red passive galaxies, and inducing a related morphological change.

Isolating the physical processes bringing about such transformations is crucial. High density environments within galaxy cluster may simultaneously affect various properties of infalling galaxies through interactions with the hot intracluster medium (ICM) or high speed galaxy-galaxy encounters. While strong tidal interactions, or harassment \citep{boquien09, moore96}, are capable of stripping outer material of galaxies, the raw material for star formation in galaxies (namely the neutral hydrogen gas) is seldomly affected. This store of neutral gas may only be dislodged when cluster galaxies pass through the ICM at high speed. Interactions with the ICM result in stripping of the cold neutral atomic gas from the star-forming disks \citep[ram-pressure stripping][]{gunngott72} or removal of hot gaseous halo through starvation, irrespective of halo mass \citep{larson80}.  

\citet[][hereafter K17]{kk17} explored the implications of such physical drivers on galaxies' intrinsic structure, revealed after removing the smooth component of the surface brightness distribution. For both cluster and field galaxies, they found that the degree of structural disturbance is better represented by the amount of light left behind in galaxy residuals, with the quantitative  asymmetry \citep{conselice00} in galaxy residuals being sensitive to the causes of structural disturbance. Furthermore, they find a higher incidence of visually smooth passive spiral galaxies in clusters as compared to the field. This result and several complementary results \citep[e.g.][]{wolf09, cantale16} thus draw our attention to the gas removal processes dominant in the cluster environment, which result in the stripping of gas without large-scale disruption in galaxy structure. However, the extent of stripping within the cluster and the timescales over which these processes occur are still open questions.

Revisiting the global picture of galaxy evolution in the context of cluster environment, radial trends are observed in various galaxy properties, like star formation \citep{vlinden10, mahajan12}, when observed as a function of distance from the cluster centre. In the cluster cores, the early-type passive galaxies are found to have lower line-of-sight velocities as compared to the late-type star-forming counterparts \citep{mahajan11}, indicating star formation evolution on a radial scale within the cluster. These observations suggest the possibility of using the projected radial information as well as the line-of-sight velocities of galaxies in clusters, to analyse the galaxy properties in a dynamical environment. This 2-D projected `phase--space' can be used to statistically segregate galaxies according to their orbital histories and time of accretion.   

Indeed, \citet{gill05} discovered that the satellite galaxy populations are better distinguished on this 2-D phase--space, as compared to just 1-D clustercentric radius. Moreover, it is also found that $\sim50\%$ galaxies within $~2R_{\rm 200}$ have already passed through the core at least once \citep{balogh00}. Works like \citet{mahajan11} and \citet{oman13} demonstrate that it is possible to trace orbital histories of galaxies by analysing 3-D phase--space using simulated galaxy clusters representing the local population. Such assembly history of galaxy clusters enables us to associate cluster galaxy populations to various regions of this phase--space. This can separate the galaxies which are infalling from the galaxies which are virialised in the cluster core, and identify galaxies which have undergone at least one pericentric passage through the core (known also as `backsplash' galaxies). 

Observationally, projection effects make it difficult to identify these backsplash galaxies on the 2-D phase--space \citep[See Figure 6 in][]{rhee17}. However, observations of e.g. the gas component of galaxies can be used to identify such galaxies, as demonstrated by \citet{yoon17} who combined the HI morphologies of galaxies in the Virgo cluster with their location on the projected phase--space to identify the most likely backsplash galaxy subpopulation. The projected phase--space, therefore, presents a powerful tool to address current questions like quenching and group pre-processing in infalling galaxies \citep[e.g][]{oman16}, and their transformation by observing them `in action' in dynamical environments \citep[e.g.][]{jaffe16}. This approach has further been utilised to study the velocity modulation of various galaxy properties like quenching timescales \citep{muzzin14,taranu14}, HI  and ionised gas stripping in galaxies \citep{hfernandez14,jaffe15,rudnick17}, and dust temperatures in galaxies \citep{noble16}. 

In this paper, we seek to explore the links between the deviations in the stellar distribution in galaxies, as introduced in K17, and the star formation history of galaxies derived from spectroscopic information. Further, focussing on the cluster galaxies, we investigate the behaviour of these structural distortions in galaxies over the dynamical cluster environment, given by the projected phase--space. On connecting these trends with the relative stellar ages of cluster galaxies, we attempt to further our knowledge regarding the observed morphological and star formation transition in galaxies as they experience the high density cluster environment. 

This paper is structured as follows: Section~\ref{secdata} presents the dataset, and describes the measures for global environment and galaxy structure. Section~\ref{sec2:sfh} and ~\ref{sec:sfh-struct} analyse and discuss the star formation history of galaxies, and corrsponding correlations with environment and quantitative structure. In Section~\ref{sec:pps} we introduce the projected phase--space, and analyse quantitative galaxy structure and stellar ages as a function of position on the phase--space. Finally, in Section~\ref{sec:conc} we present a discussion of our results and conclusions. Throughout this paper, we use the standard
$\Lambda$CDM Cosmology with $h_{0} = 0.7$, $\Omega_{\Lambda} = 0.7$
and $\Omega_{\rm m} = 0.3$, and AB magnitudes.

\section[]{Data}
\label{secdata}

The work presented in this paper employs the spectroscopic data sample of K17 for the ten high-$z$ clusters from the ESO Distant Cluster Survey (EDisCS), which is a multiwavelength survey
comprising 20 fields containing galaxy clusters in the redshift range
$0.4<z<1$ \citep{white05}.

Each of these ten high-$z$ clusters have {\em HST} $I-$band imaging using the $F814W$ filter on the ACS Wide Field Camera, and $V-$, $R-$, $I-$band photometry using the FORS2 on the {\em ESO Very Large Telescope (VLT)} \citep{white05}. The {\em HST} observations comprised a total of five pointings per
 field centred at the location of the brightest cluster galaxy such that the central region of each cluster has higher $S/N$ as compared to the surrounding area \citep{desai07}. The spectroscopy with FORS2/VLT was carried out on a subset of galaxies from the photometric sample selected using the $I$-band magnitude and the best-fit photometric redshift, as described in \citet{halliday04} and \citet{mjensen08}, which ensures an average success rate of $\sim$50\% with at most $\sim$3\% cluster members not satisfying the selection criteria. We refer the reader to K17 for in-depth details regarding the survey data products and the follow-up data, which we have not included in this paper to avoid repetition.

Our spectroscopic data sample has a mass-completeness limit of $\log M_\ast/M_\odot = 10.6$ and $S/N>2$ in the continuum \citep[which is adopted from ][]{rudnick17}. The $S/N$ lower limit ensures reliability in the measurement of spectral features which are utilised to analyse the star formation histories of galaxies (Section~\ref{sec2:sfh}). Note however that Section~ \ref{sec2:sfh} uses an extended sample with the empirical mass-completeness limit of $\log M_\ast/M_\odot = 10.4$ which permits us to directly compare our results to the spectral analysis presented in \citet{rudnick17}. Furthermore, this paper implements the structural analysis of galaxies carried out using the {\em HST} imaging as introduced in K17, and are further described in Section~\ref{subsecgalstruct}. Table~\ref{table1} summarises the properties of these clusters used in this analysis.

\begin{table*}
\caption[Summary of the  properties of the ten high-$z$ clusters from the EDisCS]{The properties of EDisCS high-$z$ cluster sample as was introduced in K17. Columns~1--6 contain the cluster ID,
 cluster redshift, cluster velocity dispersion, cluster mass \citep{finn05}, the number of spectroscopically confirmed cluster members \citep{halliday04,mjensen08}, and the virial radius in terms of $R_{\rm 200}$ \citep{finn05}.
 \label{table1}\newline}
\begin{center} 
\centering
\begin{tabular}{lccccc}
 
 \hline
 Cluster &  $z_{\rm cl}$ & $\sigma_{\rm cl}$ & $\log M_{\rm cl}$ & No.\ of spec. & $R_{\rm 200}$ \\
 &  & (km$\,$s$^{-1}$) &  ($M_\odot$) & members & (Mpc)\\
 \hline
 {\it Clusters} & & & & & \\
 cl1232$-$1250  & 0.5414 & 1080$^{+119}_{-89}$ & 15.21 & 54 & 1.99 \\
 cl1216$-$1201$^\dag$   & 0.7943 & 1018$^{+73}_{-77}$ & 15.06 & 67 & 1.64 \\
 cl1138$-$1133$^\dag$  & 0.4796 & 732$^{+72}_{-76}$ & 14.72 & 49 & 1.4 \\ 
 cl1354$-$1230$^\dag$   & 0.7620 & 648$^{+105}_{-110}$ & 14.48 & 22 & 1.05 \\
 cl1054$-$1146$^\dag$   & 0.6972 & 589$^{+78}_{-70}$ & 14.38 & 49 & 0.99 \\
 cl1227$-$1138$^\dag$  & 0.6357 & 574$^{+72}_{-75}$ & 14.36 & 22 & 1.0 \\
 cl1138$-$1133a   & 0.4548 & 542$^{+63}_{-71}$ & 14.33 & 14 & 1.05 \\
 cl1037$-$1243a$^\dag$ & 0.4252 & 537$^{+46}_{-48}$ & 14.33 & 43 & 1.05 \\
 cl1054$-$1245$^\dag$   & 0.7498 & 504$^{+113}_{-65}$ & 14.16 & 36 & 0.82 \\
 cl1227$-$1138a   & 0.5826 & 432$^{+225}_{-81}$ & 13.69 & 11 & 0.61 \\
 cl1040$-$1155$^\dag$  & 0.7043 & 418$^{+55}_{-46}$ & 13.93 & 30 & 0.7 \\
 {\it Groups} & & & & & \\
 cl1037$-$1243 & 0.5783 & 319$^{+53}_{-52}$ & 13.61 & 16 & 0.58 \\
 cl1103$-$1245a  & 0.6261 & 336$^{+36}_{-40}$ & 13.66 & 15 & 0.59 \\
 cl1103$-$1245b  & 0.7031 & 252$^{+65}_{-85}$ & 13.27 & 11 & 0.42 \\
 
 \hline

\end{tabular}
\end{center}
\footnotesize{$^\dag$ Clusters used for projected phase--space analysis in Sec~\ref{sec:pps}} 

\end{table*}

\subsection[]{Global environment: cluster membership}
\label{subsecenv}

We utilise the spectroscopic cluster membership to define the global environment for our parent spectroscopic sample.  
Broadly, a galaxy is considered a member of a cluster if its spectroscopic redshift lies within $\pm3\sigma_{\rm cl}$ from the average cluster redshift $z_{\rm cl}$ else they are considered to be in the field sample \citep{mjensen08,halliday04}. Both $z_{\rm cl}$ and $\sigma_{\rm cl}$ were computed by successive $\pm3$  sigma-clipping in the rest-frame peculiar velocity distribution of galaxies till convergence was achieved. This iterative method used the median for estimating $z_{\rm cl}$ while $\sigma_{\rm cl}$ was determined using biweight estimator, the details of which are discussed in \citet{mjensen08}. To be accepted in the parent field sample, we allow the EDisCS galaxies to lie within $\pm0.2$ from the parent cluster redshift $z_{\rm cl}$ \citep{mjensen08}. Furthermore, we avoid potential biases due to similar redshift distribution of cluster and field galaxies by selecting the field galaxies within $\pm0.05$ from the minimum and maximum $z_{\rm cl}$ from the entire cluster sample. Table~\ref{table1} also includes the secondary structures denoted by `a' or `b' following the main cluster ID \citep{white05, mjensen08}, and were further identified by \citet{poggianti09-2} into `clusters' ($\sigma_{\rm cl} > 400$ km$\,$s$^{-1}$) and `groups' ( $160 < \sigma_{\rm cl} < 400$ km$\,$s$^{-1}$; containing a minimum of 8 spectroscopic members). Our definition of global environment however,\ remains independent of the host cluster/group identification.

\subsection[]{Galaxy structure: qualitative and quantitative estimators}
\label{subsecgalstruct}

In this study we employ the extensive qualitative and quantitative
measurements of galaxy structure presented in K17. We obtain the
morphologies for the galaxies through visual classification using the {\em HST} $I-$band images carried out by five classifiers, the procedure of which is detailed in \citet{desai07}. While the {\em HST} morphology catalog includes finer morphological classes, we have combined them into four broad bins: ellipticals,
lenticulars, spirals, and irregulars.

In addition, we assess qualitative galaxy structure by classifying the degree of visual asymmetry and assigning a likely cause of disturbances in the visible galaxy structure. These included identifying signatures of galaxy interaction, tidal stripping, ongoing mergers, or any other structural asymmetry that could not be associated with any external processes. These classifications were carried out by three classifiers, and each galaxy from the HST morphology catalog was assigned a final asymmetry and disturbance index based on the weighted combination of individual classifications (for details refer to K17).

Furthermore we utilise structural information embedded in galaxy residuals after removing the bulk light by single Sersic profile fitting with \textsc{galfit} \citep{barden12}. We measure the fractional contribution to galaxy light distribution in residuals through residual flux fraction \citep[$RFF$;][]{hoyos11,hoyos12}, and define the asymmetry in galaxy residuals \citep[$A_{\rm res}$;][]{conselice00} to be the departure from a symmetric light distribution under 180 degree rotation. K17 demonstrated that these measurements of $RFF$ and $A_{\rm res}$ together are crucial in identifying structural signatures in galaxies. Individually, $RFF$ is found to be sensitive to the degree of structural disturbance or `roughness' in galaxies but is indecisive in determining the physical causes of such disturbance. The $A_{\rm res}$ however is efficient in discriminating between likely physical processes causing the structural asymmetry, and hence the disturbance in galaxy structure. Figure 5 in K17 highlights these characteristics where visually classified mergers are found to have have the highest $A_{\rm res}$ while internally asymmetric, and interacting galaxies display some of the lowest $A_{\rm res}$ values. An in-depth discussion of these measurements, and their correlation with qualitative structural disturbances and morphology is further included in K17.\\   

\section[Star formation history from spectral indices]{Star formation history from spectral indices}
\label{sec2:sfh}
In addition to the galaxy structural measurements of K17, in this paper we use the spectroscopy to derive star formation histories for individual galaxies from spectral indices.  A galaxy that has experienced a recent burst of star formation ($\sim$0.5--2 Gyr ago), or has had its star formation recently truncated shows strong Balmer absorption lines in its spectrum. This is because Balmer absorption lines, attributed to A-type stars, peak in strength after the hot O and B stars have ceased their evolution. In particular, we utilise $H$\textsc{$\delta$} 
and $H$\textsc{$\gamma$} Balmer lines as they can be robustly measured in the continuum spectra for the wavelength range of our sample \citep[for details refer to ][]{rudnick17}. 
The $H_{\delta\rm{A}}$ and $H_{\gamma\rm{A}}$ indices (incorporating the EW of the $H$\textsc{$\delta$} 
and $H$\textsc{$\gamma$} lines), therefore are indicative of the time since the recent episode of star formation in the galaxies \citep{worthey97}. Moreover, these indices also depend on metallicity and $\alpha/Fe$ ratio, which is important for older stellar populations \citep{thomas04}. A measure of the overall age of the stellar population, on the other hand, could be represented by the strength of the break occurring at {4000}\AA{} \citep[$D_{n}4000$ hereafter]{balogh99} as it traces recent star formation on timescales longer than those indicated by Balmer absorption. This break depends on metallicity and 
arises due to the accumulation of metal absorption lines on the blue side of the spectrum. \citet{kauffmann03} shows that the locus of the plane constructed using these spectral indices 
give a good relative (though not unique) measure of the star formation history of galaxies. Moreover, the variation in $H_{\delta\rm{A}}$, $H_{\gamma\rm{A}}$, and $D_{n}4000$ is relatively 
insensitive to dust extinction, making them a reliable diagnostic to assess whether star formation occurred in bursts or in continuous mode.

In this paper, we present a simplified stellar population analysis which will be detailed in Deger et al. (2018b, in prep). 
The relative star formation history (SFH hereafter) of galaxies in the parent spectroscopic sample was determined through the indices obtained using the spectral decomposition of the galaxies and requiring empirical mass completeness of $\log M_\ast/M_\odot > 10.4$, and  $S/N>2$ in the continuum \citep{rudnick17}. Although the detailed decomposition of galaxies' spectrum is performed as a part of the work presented in Rudnick et al., the analysis pertaining to studying the interlink between SFH and galaxy structure adopts the above a mass-completeness limit. Specifically, we use the $H_{\delta\rm{A}}$ and $H_{\gamma\rm{A}}$ indices, and $D_{n}4000$ in the continuum of the spectra to investigate the SFH of galaxies. We further improve the S/N of Balmer absorption by taking the average of $H_{\delta\rm{A}}$ and $H_{\gamma\rm{A}}$ indices which can also mitigate the systematic offsets that each index has when compared to models \citep{rudnick17}.

\begin{figure}
 \centering
 \includegraphics[width=0.5\textwidth]{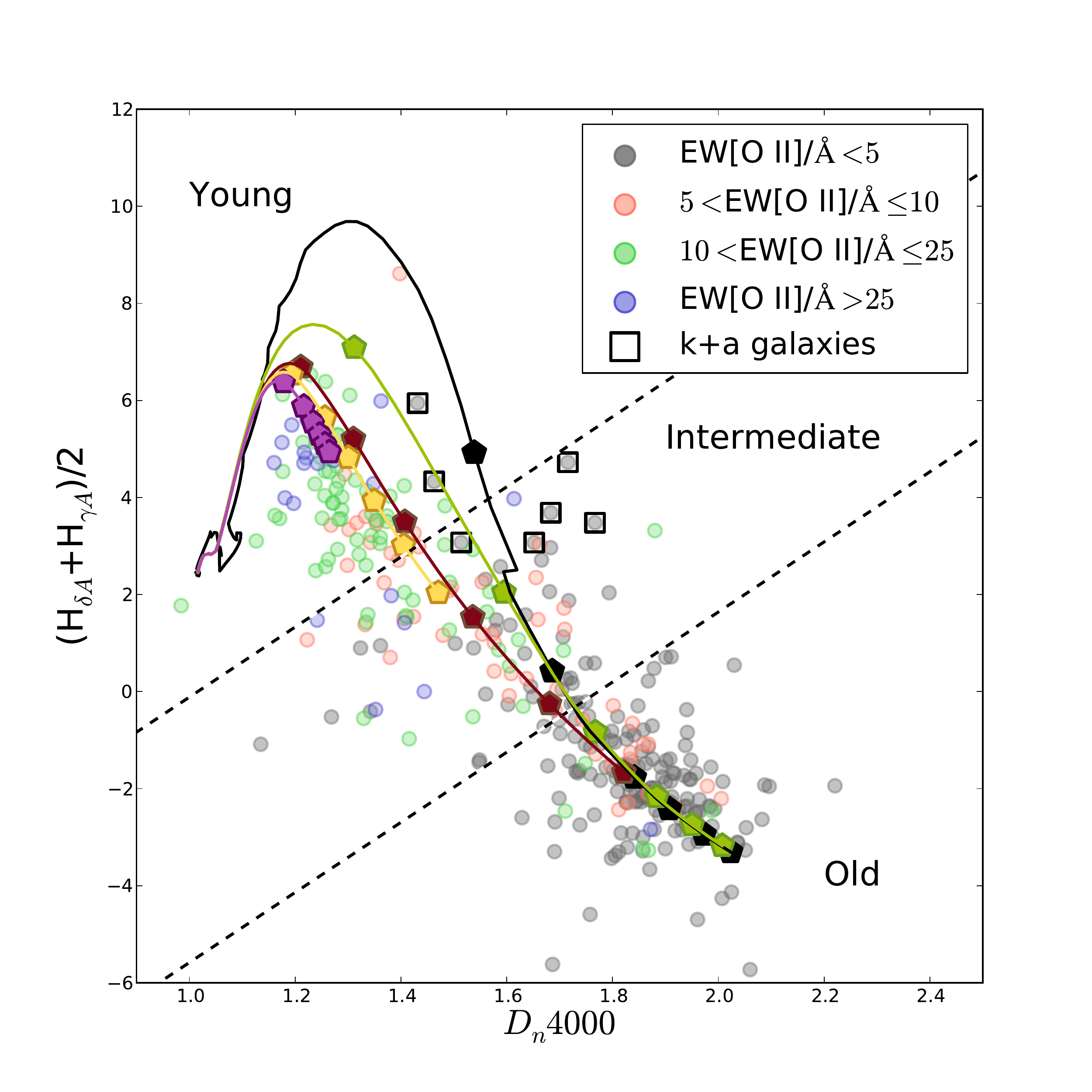}
 \caption[The ($H_{\delta\rm{A}}$+$H_{\gamma\rm{A}}$)/2 vs $D_{n}4000$ plane: Star formation history of EDisCS galaxies]{\label{dak_b_model} The ($H_{\delta\rm{A}}$+$H_{\gamma\rm{A}}$)/2 vs 
 $D_{n}4000$ plane for all the galaxies with $\log M_\ast/M_\odot > 10.4$ in the EDisCS spectroscopic sample, irrespective of environment. The colour of the points indicate the EW[O\textsc{ii}], an indicator of star formation activity. The galaxies are further divided into galaxies with young, intermediate and older stellar populations, as described in the text (black dashed lines). The overplotted tracks denote the modelled evolution of stellar population with solar metallicities \citep{bruzual03} formed in a single burst (black), displaying constant star formation (magenta), and the three exponentially declining SFH with timescales of 0.3 Gyr(green), 1 Gyr (red) and 2 Gyr (yellow). All the points on these model tracks are spaced at 1 Gyr intervals increasing to lower right. In addition, the `k+a' galaxies in the sample are represented by black squares.}   
\end{figure}

Figure \ref{dak_b_model} shows the ($H_{\delta\rm{A}}$+$H_{\gamma\rm{A}}$)/2 vs $D_{n}4000$ for the all 
the galaxies in the sample, irrespective of environment, with the mass completeness limit of $\log M_\ast/M_\odot = 10.4$. The points are colour-coded by the [O\textsc{ii}]$\lambda3727$ EW which traces the current star formation in galaxies \citep{gallagher89,kennicutt92}. Further, to understand the distribution of points on this plane, we use the model tracks with solar metallicity derived from the \citet{bruzual03} stellar population synthesis code. These model tracks include a single burst model of star formation, a constant star formation history model, and three models with exponentially declining SFHs with a timescale of 0.3, 1, and 2 Gyr. All the models extend to an age of 6 Gyr, which roughly corresponds to star-forming at $z=2.5$ for a galaxy observed at the lower end of the EDisCS redshift range $z=0.4$. 

Following from Figure \ref{dak_b_model}, galaxies are separated based on their relative stellar ages as galaxies with - `young' (top left), `intermediate' (middle), and `old' (bottom right) stellar populations. The line dividing the galaxies with older stellar populations was determined based on the isolation of the peak of these galaxies in the histogram of the galaxy sample \citep[Figure 2,][]{rudnick17}. It is thus observed that galaxies selected in such a way have negligible recent star formation ($\sim 6$ e-foldings of SFH). The rest of the galaxies were then split into galaxies with intermediate and young stellar populations. The division for these galaxies was chosen to split the difference in the Balmer absorption and $D_{n}4000$ strength between the exponential SFH with a timescale of 2 Gyr and a constant SFR model. A detailed description of these borders is presented in \citet{rudnick17}.  

Figure~\ref{dak_b_model} also shows the post-starburst or `k+a' galaxies in the sample. The relative fractions of `k+a' galaxies in the EDisCS clusters were previously studied by \citet{poggianti09-2}, who use Gaussian line fitting technique to determine the EWs of Balmer absorption. In the present work, the `k+a' galaxies are defined using the more accurate Balmer absorption from $H_{\delta\rm{A}}$ and $H_{\gamma\rm{A}}$ indices. The specific selection criteria we use are EW[O\textsc{ii}]$ <$ 5\AA{}, and ($H_{\delta\rm{A}}$+$H_{\gamma\rm{A}}$)/2$>$3\AA{}. The sample of galaxies thus identified as `k+a' are all cluster members (except one), and is different from \citet{poggianti09-2}, primarily because their method was different. These galaxies will be further investigated in Section~\ref{sec:sfh-struct}.

\section[Correlation of star formation history with galaxy structure and environment]{Correlation of star formation history with galaxy structure and environment}  
\label{sec:sfh-struct}

The results of K17 show a strong correlation between quantitative measurements of structural disturbances ($RFF$ and $A_{\rm res}$) and qualitative measurements of visual structural disturbances and morphology. Moreover, the analysis of the $(U-V)$ vs. $(V-J)$ colour plane reveals that passive galaxies are mostly structurally smoother and more symmetric, with low values of $RFF$ and $A_{\rm res}$. This link between galaxy `roughness', `asymmetry', and current star formation, indicates a possible correlation of intrinsic galaxy structure with the SFH of galaxies. 

\begin{figure*} 
  \centering
  \includegraphics[width=1\textwidth]{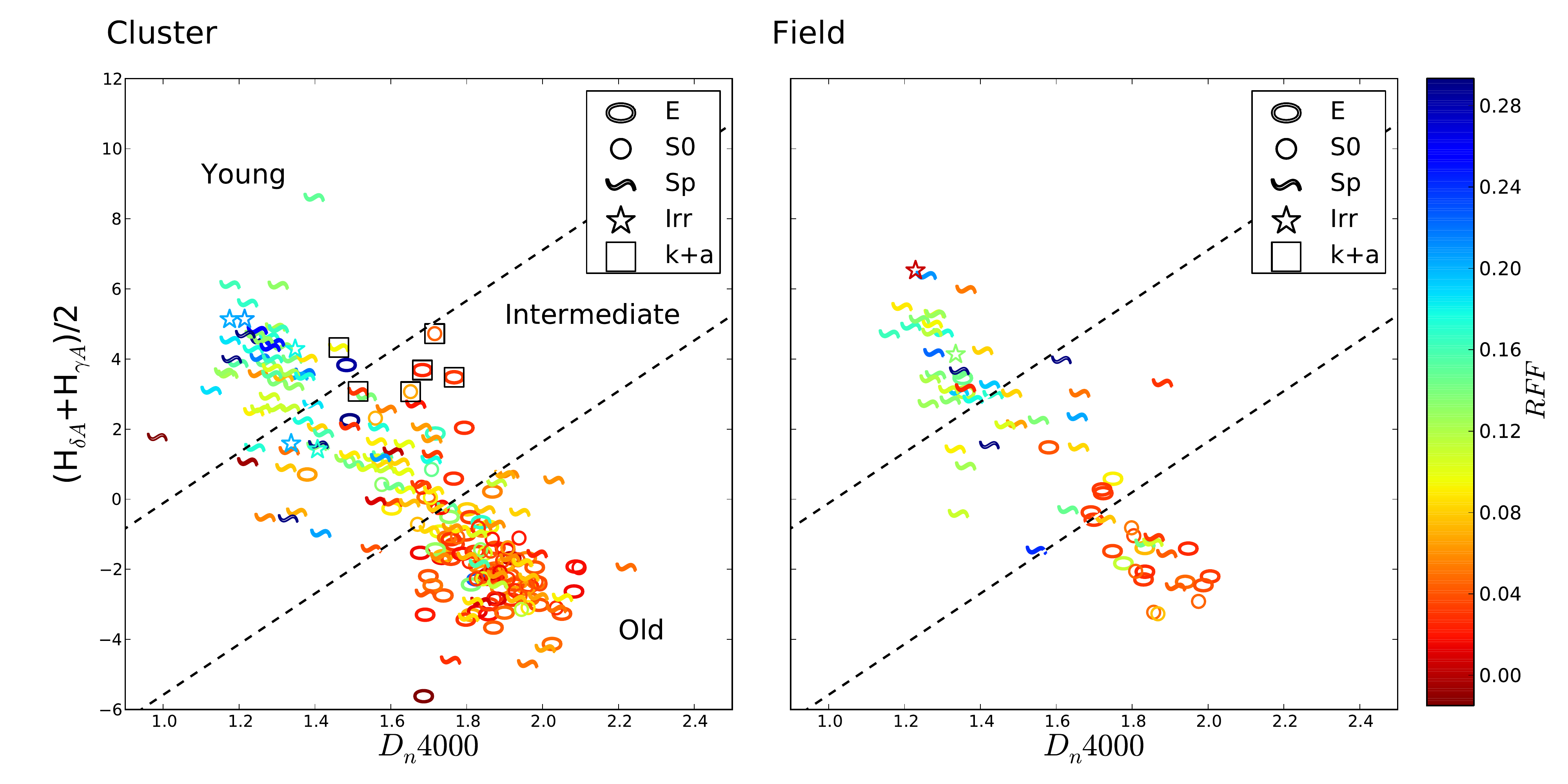}
 
  \caption[The ($H_{\delta\rm{A}}$ + $H_{\gamma\rm{A}}$)/2 vs $D_{n}4000$ plane colour-coded according to $RFF$ for the cluster and field galaxies]{\label{d4k_babs_rff} The ($H_{\delta\rm{A}}$ + $H_{\gamma\rm{A}}$)/2 vs $D_{n}4000$ plane is colour-coded according to
   $RFF$ values for the cluster (left panel) and field (right panel)
   galaxies: redder colours indicate smoother, undisturbed galaxies
   while bluer colours denote `roughness'. The symbols indicate the morphology of galaxies, as explained in Section~\ref{secdata}. As in Figure \ref{dak_b_model},
   the dashed lines show the selection boundaries for galaxies with young, intermediate and older stellar populations. Galaxies with ongoing star formation or with younger stellar populations are consistently found to be `rougher' irrespective of morphology or global environment.}
  \end{figure*}

\begin{figure*} 
  \centering
  \includegraphics[width=1\textwidth]{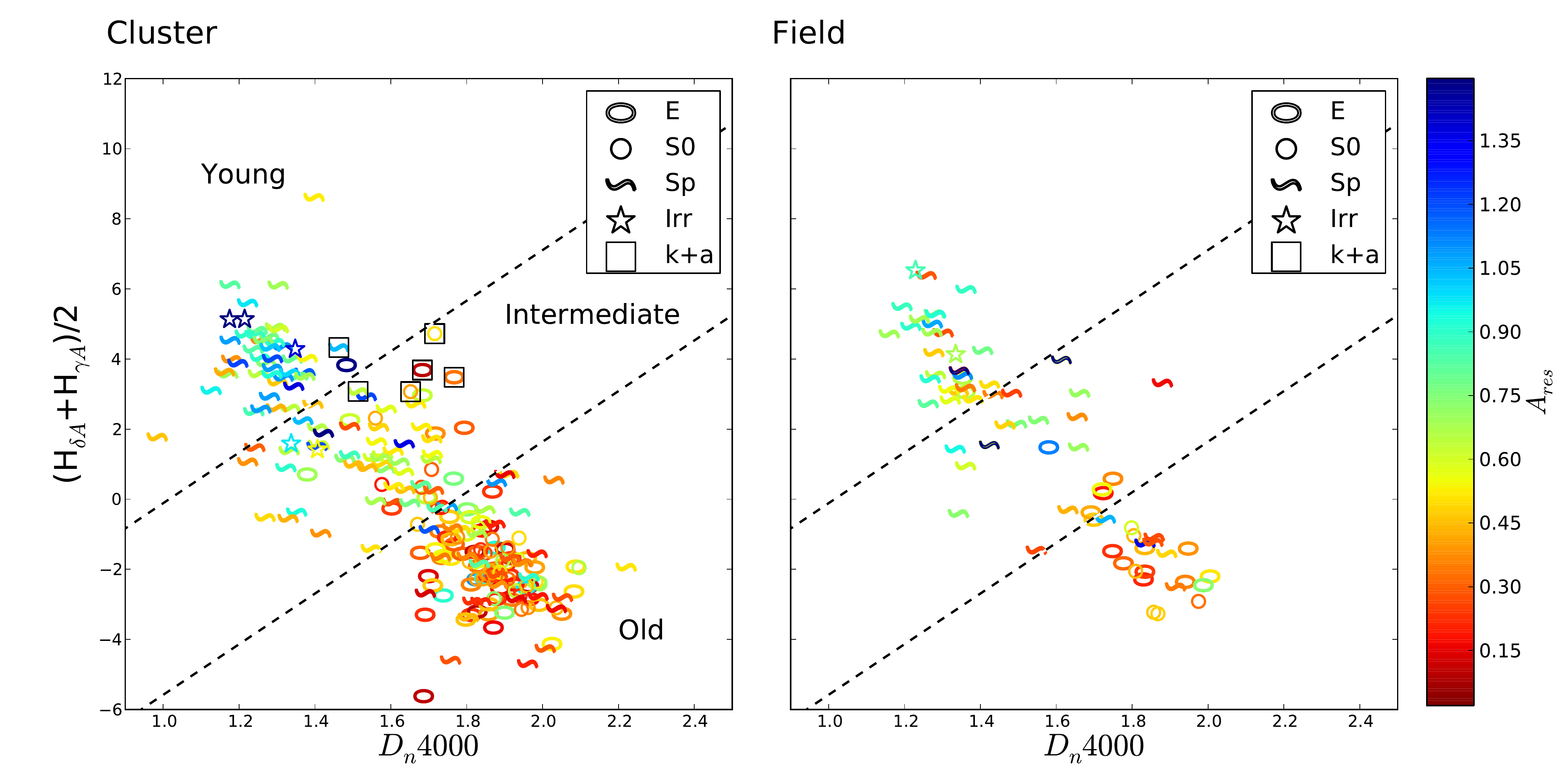}
 
  \caption[The ($H_{\delta\rm{A}}$ + $H_{\gamma\rm{A}}$)/2 vs $D_{n}4000$ plane colour-coded according to $A_{\rm res}$ for the cluster and field galaxies]{\label{d4k_babs_ares} Similar to Figure \ref{d4k_babs_rff}, the ($H_{\delta\rm{A}}$ + $H_{\gamma\rm{A}}$)/2 vs $D_{n}4000$ plane is
  colour-coded according to $A_{\rm res}$ for cluster (left
   panel) and field (right panel) galaxies: redder colours denote `symmetric' galaxies with low $A_{\rm res}$
   while bluer colours show galaxies with higher asymmetry. Galaxies with ongoing star formation or with younger stellar populations appear to be structurally asymmetric irrespective of morphology or global environment.}
  \end{figure*}

We therefore investigate the variations in $RFF$ and $A_{\rm res}$ with the evolution of star formation over a global cluster and field environment. We use the cluster sample and the redshift-controlled field sample for this analysis, as described in Section~\ref{subsecenv}. Figures~\ref{d4k_babs_rff} \& ~\ref{d4k_babs_ares} show the average Balmer absorption plotted against $D_{n}4000$ for the spectroscopic sample and colour-coded with $RFF$ and $A_{\rm res}$. At a glance, the $RFF$ and $A_{\rm res}$ display a strong interdependence with the SFH of galaxies indicated by the average age of their stellar population. Indeed, galaxies with recent star formation are asymmetric and show a higher degree of roughness/texture while galaxies with high $D_{n}4000$ display smoother structure with low $RFF$ and $A_{\rm res}$ values. This observation is further strengthened by the distribution of galaxy morphologies on this plane. As expected, galaxies displaying early-type morphologies populate the lower right region with high $D_{n}4000$, while the late-type/irregular galaxies show ongoing star formation as indicated by relatively higher average Balmer absorption and [O\textsc{ii}] emission (cf. Figure~\ref{dak_b_model}). Comparing both panels of Figures ~\ref{d4k_babs_rff} \& ~\ref{d4k_babs_ares}, the only environmental effect seen is the build-up of older population in clusters. 
This implies that, irrespective of morphology or global environment, it is possible to broadly identify galaxies with different SFHs based on the roughness and asymmetry in structure. This result has important connotations as both $RFF$ and $A_{\rm res}$ are purely structural quantities measured from the deviations in the galaxy light from a smooth symmetric profile.

Revisiting the `k+a' galaxies as described in Figure~\ref{dak_b_model}, it is worth noting that all the `k+a' galaxies in the sample are detected in clusters\footnote{The only field `k+a' galaxy is not contained in the redshift-controlled field sample. } and lie in the region which has high $RFF$ and $A_{\rm res}$ on the average Balmer absorption vs $D_{n}4000$ plane. Moreover, the $RFF$ and $A_{\rm res}$ of these `k+a' galaxies have an average value of $0.05$ and $0.5$ respectively, which is on the border of the definition separating the rough/smooth and symmetric/asymmetric galaxies in the sample (cf. Section~\ref{sec:pps-sfh}). This result may support the findings of \citet{pawlik16} which show that the shape asymmetry of galaxies, a similar but slightly different measure of quantitative morphology, is found to be strongly correlated with the age of starburst activity in galaxies.  They find young starbursts displaying distinct asymmetric features indicative of recent merger activity. However, our sub-sample is too small to robustly confirm this result with the measurement of $RFF$ and $A_{\rm res}$. A more thourough analysis of these `k+a' galaxies will be presented in Deger et al. (2018b, in prep).

  \clearpage
\begin{landscape}
  \begin{figure}
  \hspace{-2.2cm}
  \includegraphics[width=1.5\textwidth]{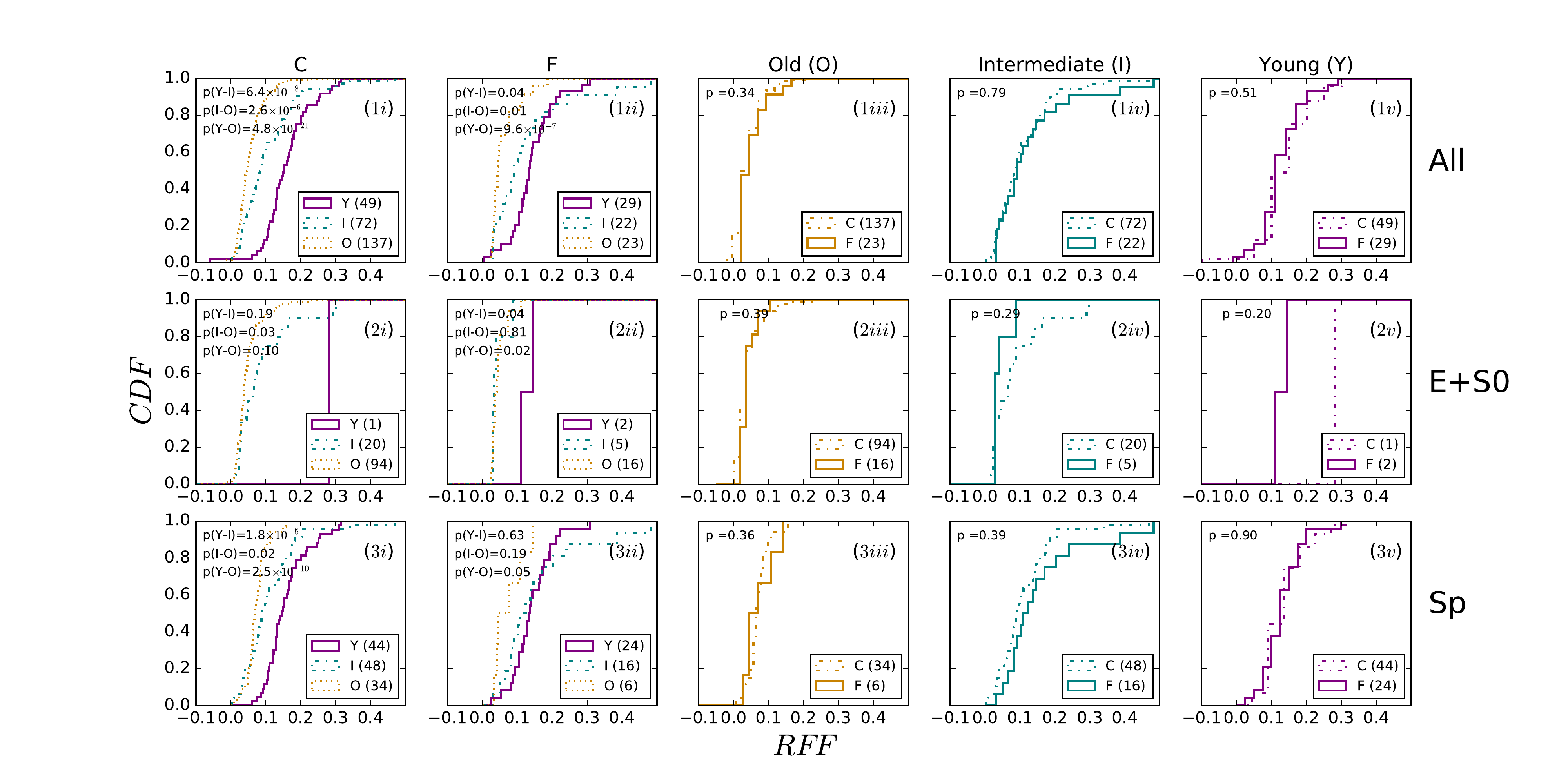}
 
  \caption[$RFF$ distribution for the galaxies with Old (O), Intermediate (I) and Young (Y) stellar ages in Cluster (C)/Field (F) environment]{\label{cdf_r} $RFF$ distribution for the galaxies with Old (O), Intermediate (I) and Young (Y) stellar ages in Cluster (C)/Field (F) environment. The galaxies are separated in different stellar ages according to the boundaries represented in Figure~\ref{dak_b_model}. Each row shows the $RFF$ distribution for the galaxies at fixed morphology: all (top), E/S0 (middle), Sp (bottom). Columns (i) and (ii) compare the  $RFF$ distribution for the three stellar ages in the cluster and field environments, respectively. Columns (iii) to (v) show the  $RFF$ distribution for cluster and field galaxies for each of the  stellar age classifications. The inset $p$-values represent the two sample K--S test probalities for the two distributions being statistically similar. It is evident that the older stellar ages display lower $RFF$ values as compared to the intermediate and younger stellar ages, and this effect is more significant in cluster environment. Further, the $RFF$ distibution for galaxies with younger, intermdeiate and older stellar populations are similar irrespective of their incidence in the cluster or field environments. }
  \end{figure}
\end{landscape}
   \clearpage

\begin{landscape}
  \begin{figure} 
  \hspace{-2.2cm}
  \includegraphics[width=1.5\textwidth]{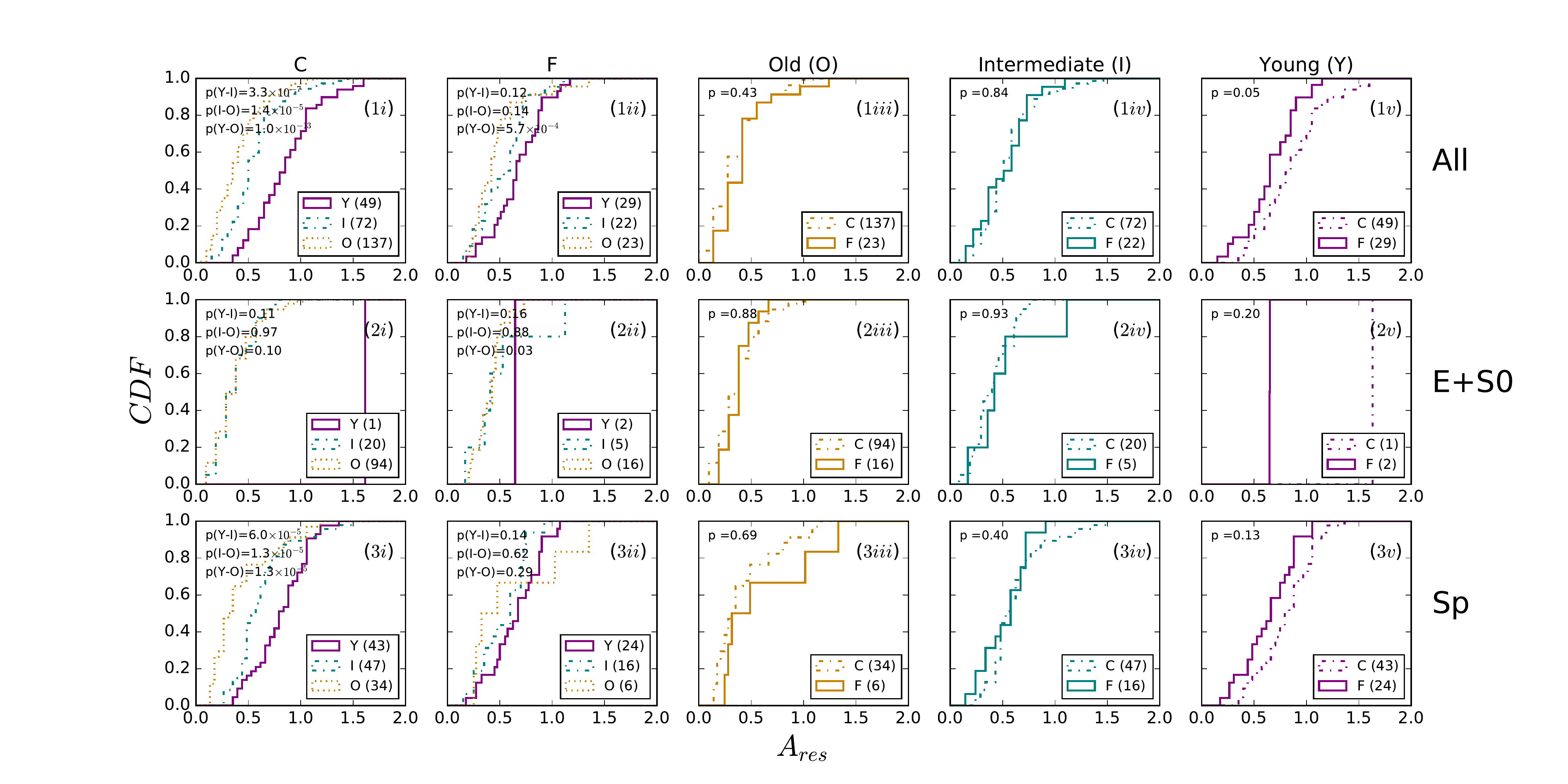}
 
  \caption[$A_{\rm res}$ distribution for the galaxies with Old (O), Intermediate (I) and Young (Y) stellar ages in Cluster (C)/Field (F) environment]{\label{cdf_a} Similar to Figure~\ref{cdf_r}, $A_{\rm res}$ distribution for the galaxies with old, intermediate and young stellar ages in cluster and field environment, at fixed morphologies. Comparing columns (i) and (ii), it is evident that the older stellar ages display lower $A_{\rm res}$ values as compared to the intermediate and younger stellar ages, and this effect is stronger in cluster environment. Further, the $A_{\rm res}$ distibution for galaxies with younger, intermdeiate and older stellar populations are similar irrespective of their incidence in the cluster or field environments.}
  \end{figure}
\end{landscape}
   \clearpage

To test how well the galaxies can be separated based on their quantitative structure, we evaluate the distributions of $RFF$ and $A_{\rm res}$  for each galaxy subpopulation defined by their stellar ages. Figures~\ref{cdf_r} and \ref{cdf_a} show the distribution of $RFF$ and $A_{\rm res}$ for galaxies with old, intermediate and young stellar ages across cluster and field environments. We use the two-sample Kolmogorov-Smirnov (K--S) tests to check the significance of the variation in the $RFF$ and $A_{\rm res}$ distributions for different stellar ages, under the null hypothesis that each sample of cluster and field galaxies with different ages are drawn from tha same $RFF$ and $A_{\rm res}$ distribution. 

Panels (1i) and (1ii) in both figures show a strong stratification of $RFF$ and $A_{\rm res}$ across stellar ages in the cluster environment ($\gg 3\sigma$ significance). However, this difference is less significant in the field galaxy population due to the smaller sample size. The morphology dependence of these two structural diagnostics is taken into account by evaluating the cumulative distribution function (CDF) at fixed morphology (Figures~\ref{cdf_r} and \ref{cdf_a}, Panels 2i, 2ii, 3i, 3ii; see also Figure 4 from K17). `Old' E+S0 are smoother as compared to `intermediate' E+S0 in the cluster environment. Further, as expected the `young' E/S0 are in the minority (only 1) in clusters and show high $RFF$ and $A_{\rm res}$. In the field environment, however, this difference becomes less significant, but the numbers are very small. This difference is further emphasised for spiral galaxies in clusters. The `old' cluster spirals are found to be smoother, and more symmetric than the `intermediate' and `young' cluster spirals, with the difference not being observed in the field. Combining the results from Figures~\ref{d4k_babs_rff} and ~\ref{d4k_babs_ares}, this effect is attributed to the characteristic morphology content observed in clusters, with high-density environment harbouring early-type morphologies and older stellar populations.

Looking at all the galaxies together, at fixed morphology, there is little to no significant difference in the distribution of $RFF$ and $A_{\rm res}$ for old, intermediate and young galaxies living in clusters or in the field. This is shown in columns (iii) to (v), which present the $RFF$ and $A_{\rm res}$ distribution for the cluster and field galaxies, categorised by their stellar ages. At fixed ages and morphologies, cluster and field galaxies have similar structural parameters. 
     
In summary, the roughness and asymmetry in the structure of a galaxy can generally be considered as the main determinants of the luminosity-weighted age of its stellar population.

There is no significant environmental difference in the distribution of $RFF$ and $A_{\rm res}$ for galaxies with similar ages and morphologies. Environment changes the proportion of `old' galaxies but does not alter the SFH and structure of galaxies, corroborating the results from K17 \citep[See also,][]{thomas10}.

\section[The Projected Phase--Space (PPS) of EDisCS clusters]{The Projected Phase--Space of EDisCS clusters}  
\label{sec:pps}

\subsection[Construction of the stacked cluster ]{Construction of the stacked cluster}  
\label{ssec:pps-construct}

The projected phase--space (PPS hereafter) is constructed using derived cluster properties $\sigma_{\rm cl}$, $R_{\rm 200}$ and cluster centre, while assuming spherical symmetry in the clusters. However, as the number of spectroscopic cluster members from each EDisCS cluster is modest, it is not possible to determine trends of galaxy properties on the PPS of each individual cluster. We therefore stack individual scaled clusters together to create a combined projected phase--space using the line-of-sight velocity and the clustercentric distance of the cluster galaxies. This method is based on the assumption of the universal mass profile for cosmological halos \citep{navarro97} and is widely used in constructing projected phase--space of clusters \citep[for example,][]{raines03}.
We select clusters for this analysis, according to the following criteria:
\begin{itemize}
 \item clusters for which we have robust spectroscopy and {\em HST} $I-$band imaging data (refer to Section~\ref{secdata}).
 \item clusters for which the distribution of the radial velocities in the cluster rest-frame is reasonably Gaussian, as determined by visual inspection.

 \item clusters that have reasonably symmetric spatial distribution of galaxies (also determined visually) in the sky with the brightest cluster galaxy located as centrally as possible.
 \item clusters that have spectroscopic coverage to at least $\sim 0.7 R_{\rm 200} $, and have 20 or more spectroscopically confirmed cluster members.\footnote{Note that not all the clusters used in the phase—-space analysis have spectroscopic coverage reaching beyond $0.7 R_{\rm 200}$.}
\end{itemize}
The reader is referred to \citet{mjensen08} for the detailed spectroscopic, spatial and velocity information on which these selections are based.
 
\begin{figure}
 
 \centering
 \includegraphics[width=0.5\textwidth]{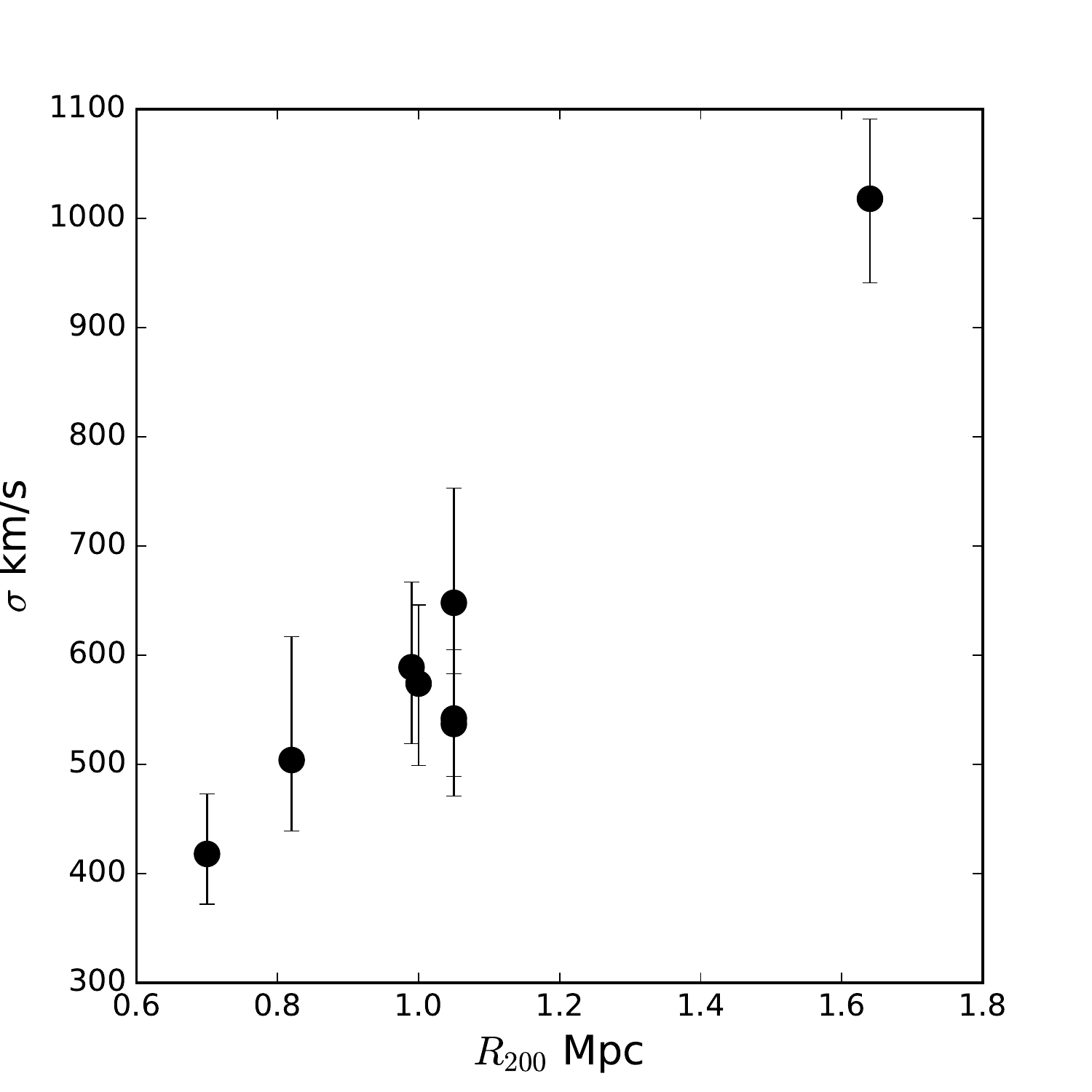}
 \caption[The distribution of $R_{\rm 200}$ and $\sigma$ for the eight clusters/secondary clusters]{\label{rsigma} The distribution of $R_{\rm 200}$ and $\sigma$ for the new cluster sample comprising eight clusters/secondary clusters. The selection criteria for these clusters is discussed in Section~\ref{ssec:pps-construct}.}   
\end{figure}

With these criteria, the new cluster sample comprises eight clusters (incuding secondary clusters) out of the original sample (cf. Table~\ref{table1}). This stacked cluster has a mean redshift $\langle{z}\rangle\simeq0.65$ and $\langle{R_{\rm 200}}\rangle\simeq1.14$ Mpc. Note that this stacking is done after scaling the cluster-centric radius of galaxies by the $R_{\rm 200}$ of the parent cluster, and the peculiar velocities by the velocity dispersion $\sigma$ of the parent cluster (cf. Table~\ref{table1}). Figure~\ref{rsigma} shows $R_{\rm 200}$ vs. $\sigma$ for the clusters used in the stacking. The final cluster sample for this analysis thus contains a total of 181 cluster galaxies. The complementary field sample is consistent with the sample presented in K17 and is described in Section~\ref{subsecenv}. Note that all the galaxies in this analysis have $\log M_\ast/M_\odot > 10.6$, which is the mass completeness limit for the parent spectroscopic sample \citep{vulcani10}.  

\begin{figure}
 \hspace{-0.7cm}
 
 \includegraphics[width=0.55\textwidth]{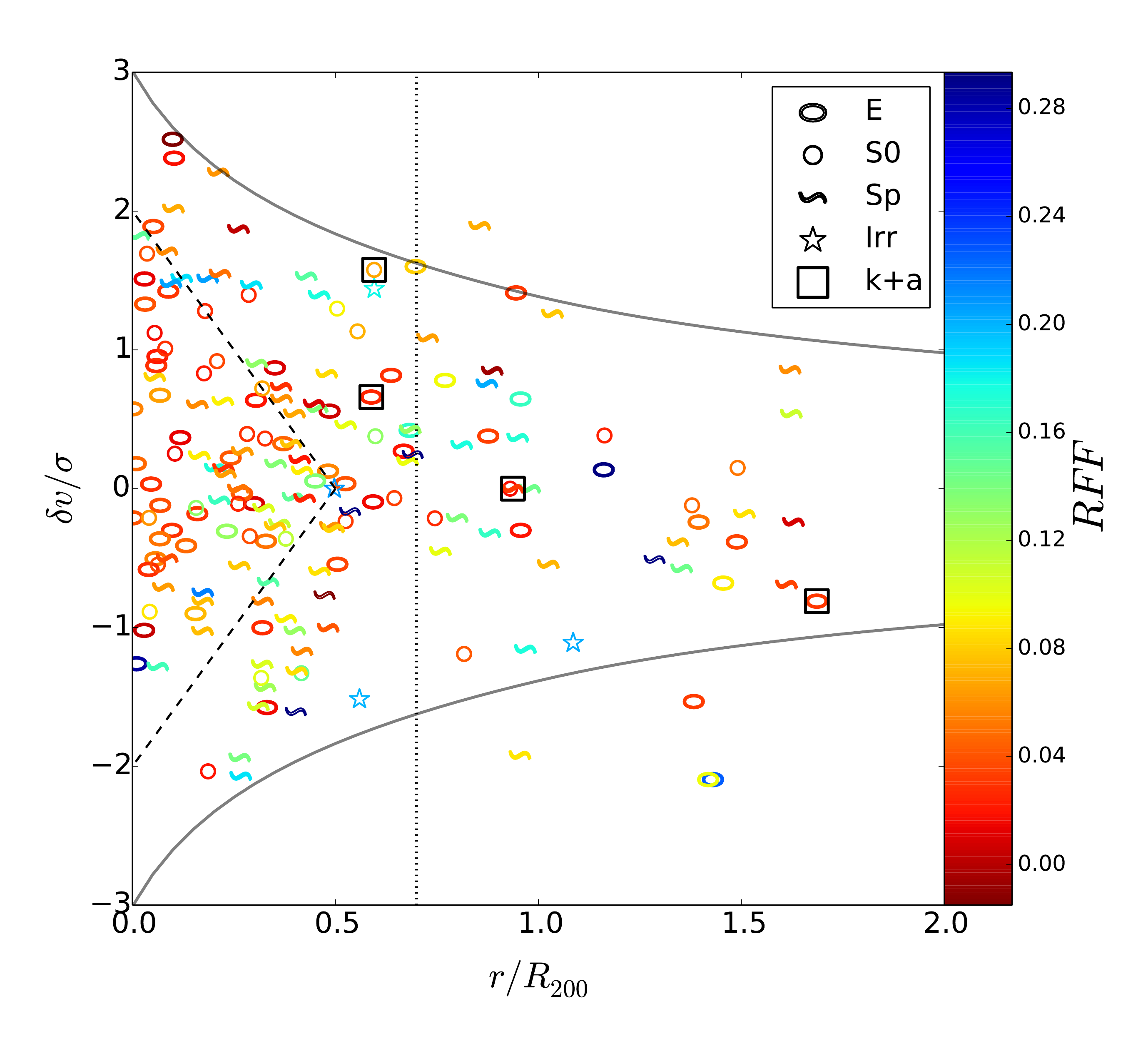}
 \caption[The projected phase--space (PPS) for EDisCS cluster galaxies with different morphologies, colour-coded according to $RFF$]{\label{pps_rff} The observed PPS for the EDisCS cluster galaxies with different morphologies, colour-coded according to $RFF$. The dashed boundary encompasses the `virial' region, as discussed in Section~\ref{ssec:pps-construct}. All the galaxies outside this region are considered as infalling galaxies. The vertical dotted line indicates the required minimum spectroscopic coverage for the stacked PPS. The solid curves are derived from the dark matter halo profile \citep{navarro97} computed using the median cluster mass of our sample, and the corresponding $\sigma_{\rm cl}$ and $R_{\rm 200}$ (Table~\ref{table1}). This observed PPS corroborates the observed distribution of galaxy morphology in galaxy clusters: the majority of the early-type galaxies lie within the virial region while late-type galaxies are prevalent in the cluster outskirts. }   
\end{figure}

\citet{rhodes13-phd} analysed the 3-D and projected phase--space for intermediate-$z$ clusters matched with the EDisCS cluster sample from N-body simulations populated by semi-analytic models, and used the co-ordinates of galaxies on the phase--space to identify various regions out to $\sim5R_{\rm 200}$. It was found from this analysis that there is $>80\%$ probability of the cluster galaxy to be bound in the cluster potential within $0.5 R_{\rm 200}$ at intermediate redshifts. Moreover, as the cluster assembles, infalling galaxies approach the cluster at higher radii while gaining velocity as they reach the cluster core. These galaxies then settle in the cluster due to dynamical friction. This process leaves a signature on the PPS, approximated best with a triangular shape \citep{mahajan11, jaffe15}. Thus, combining the findings from \citet{mahajan11} and \citet{rhodes13-phd}, we define a triangular region with the boundaries at $r = 0.5R_{\rm 200}$ and $|\delta v| = 2\sigma_{\rm cl}$ on the stacked PPS to delineate the likely virialised region. This region is then used to select the `virialised' galaxies from the `infalling' galaxies at higher $|\delta v|$ and higher $r$. 

\begin{figure}
 \hspace{-0.7cm}
 
 \includegraphics[width=0.55\textwidth]{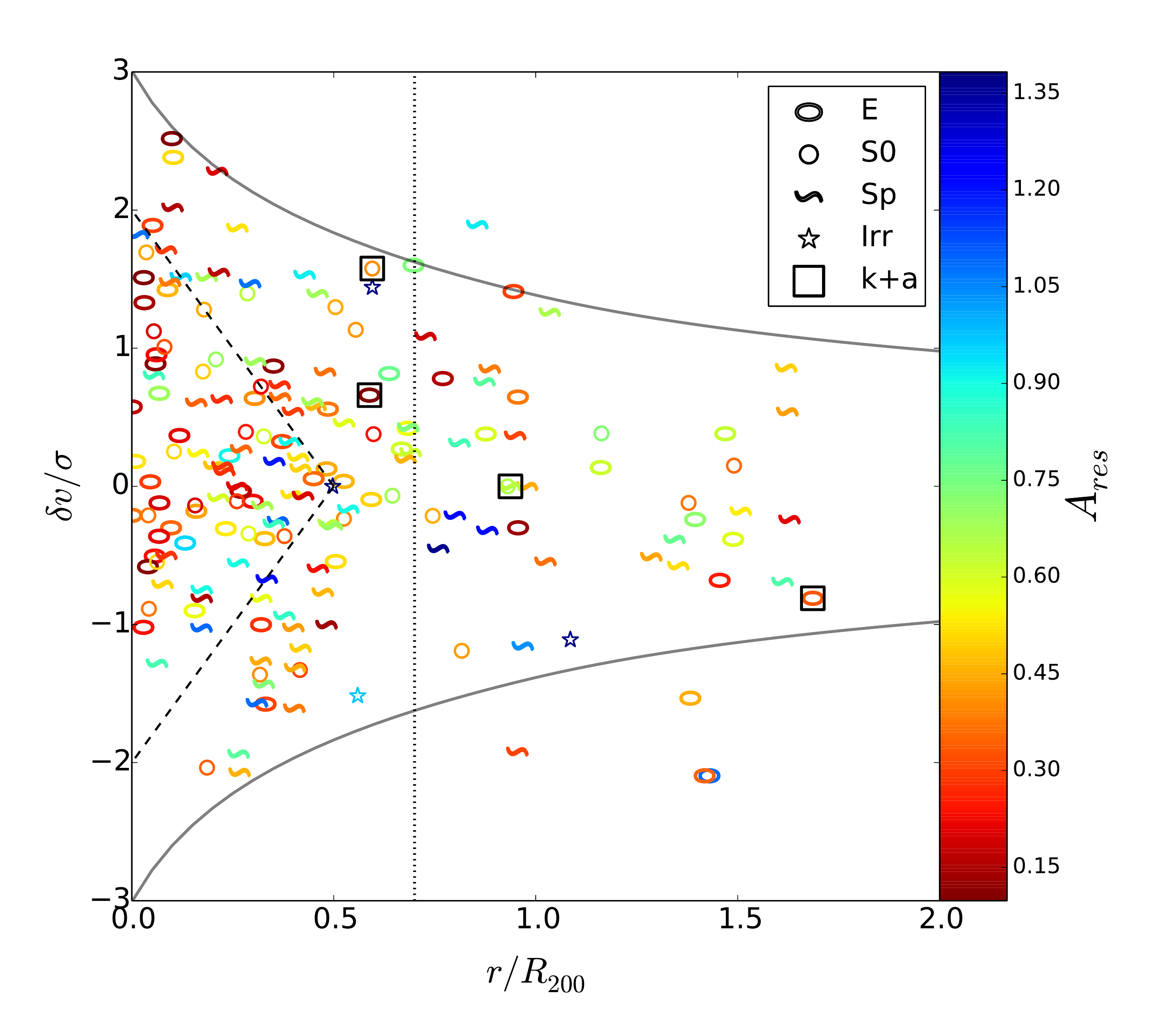}
 \caption[The PPS for EDisCS cluster galaxies with different morphologies, colour-coded according to $A_{\rm res}$]{\label{pps_ares} Similar to Figure~\ref{pps_rff}, the observed PPS for the EDisCS cluster galaxies, colour-coded according to $A_{\rm res}$.}   
\end{figure}

Figures ~\ref{pps_rff} and ~\ref{pps_ares} show the PPS for the EDisCS cluster galaxies with different morphologies, colour-coded with $RFF$ and $A_{\rm res}$ respectively. This PPS corroborates the observed distribution of galaxy morphology in galaxy clusters, with Table~\ref{table2} showing the fraction of galaxies lying within and outside the virial region for each morphology type. We use the \citet{wilson27} binomial confidence
interval to compute the $1\sigma$ uncertainty $\delta{\acute{f}_{i}}$ in the fractions
$\acute{f_{i}}$ 
\begin{equation}
\label{Equation: Frequency error}
\acute{f}_{i} \pm \delta{\acute{f}_{i}} = \frac {N_{i} + \kappa^2/{2}}{N_{\rm tot} + \kappa^2} \pm 
\frac{\kappa\sqrt{N_{\rm tot}}}{N_{\rm tot} + \kappa^2}\sqrt{f_{i}(1- f_{i}) + \frac{\kappa^2}{4N_{\rm tot}}},
\end{equation}
where $f_{i} = N_{i}/N_{\rm tot}$, and $\kappa$ is the $100(1-\alpha/2)\rm{th}$ percentile of a standard normal distribution ($\alpha$ being
the error percentile corresponding to the $1\sigma$ level (refer also to \citealt{brown01}). Note that even if $ N_{i}=0$, the estimated value of $\acute{f}_{i}$ is not necessarily $0$.

It is interesting to note that although the majority of the late-type Sp galaxies are prevalent in the cluster outskirts, the early-type E/S0 galaxies show similar fractions in both cluster regions but have lower velocity dispersion, which is an indication that they are an older cluster population. The presence of E/S0s at large clustercentric distances is partly a result 
of the known effect of pre-processing in groups prior infall into the cluster \citep{jaffe16, oman16}, or simply they all could be backsplash galaxies \citep[See also ][]{balogh00}. 
Likewise, irregulars caused by mergers are more likely to be formed outside the cluster cores,
where the relative velocities between the galaxies allow for galaxy-galaxy encounters to take place. However, the smaller numbers of irregulars in our sample prevent us from exploring this scenario further. The $RFF$ and $A_{\rm res}$ however do not seem to show a significant trend across the PPS. Moreover the `k+a' galaxies in Figures~\ref{pps_rff} and ~\ref{pps_ares} tend to avoid the cluster core, supporting the idea that they are a transitory cluster population (see also \citealt{muzzin14}).

\begin{table}
  \begin{center}
  
   \centering
   \caption[The relative fractions for galaxies classified in different cluster regions for fixed morphology]{\label{table2} The relative fractions of galaxies (all cluster members) in the virialized and infall regions denoted in Figures ~\ref{pps_rff} and ~\ref{pps_ares}, for fixed morphology. The errors on fractions denote the $1\sigma$ uncertainty computed over Wilson interval (cf. Section~\ref{ssec:pps-construct}).  \newline}
    \begin{tabular}{lcc}

    \centering
    \textbf{Morphology} & \textbf{Virial region} & \textbf{Infall region} \\
  
     & $r \lesssim 0.5R_{\rm 200}$ & $r > 0.5R_{\rm 200}$  \\
     & $|\delta v| \lesssim 2\sigma_{\rm cl}$& $|\delta v| > 2\sigma_{\rm cl}$ \\
    \hline
    \hline
    \multicolumn{3}{c}{}\\
    Ellipticals (E) & 0.52$\pm$0.07 & 0.48$\pm$0.07  \\ 
    
    Lenticulars (S0) & 0.50$\pm$0.09 & 0.50$\pm$0.09  \\ 
    Spirals (Sp) & 0.32$\pm$0.05 & 0.68$\pm$0.05\\ 
    Irregulars (Irr) & 0.30$\pm$0.20 & 0.70$\pm$0.20\\ 
    \multicolumn{3}{c}{}\\
    \hline
    \end{tabular}
  \end{center}
\end{table}

\section[Galaxy structure and star formation history as a function of location on the PPS]{Galaxy structure and star formation history as a function of location on the PPS}  
\label{sec:pps-sfh}

\begin{figure*}
 
 \hspace{-1.2cm}
 \includegraphics[width=1\textwidth]{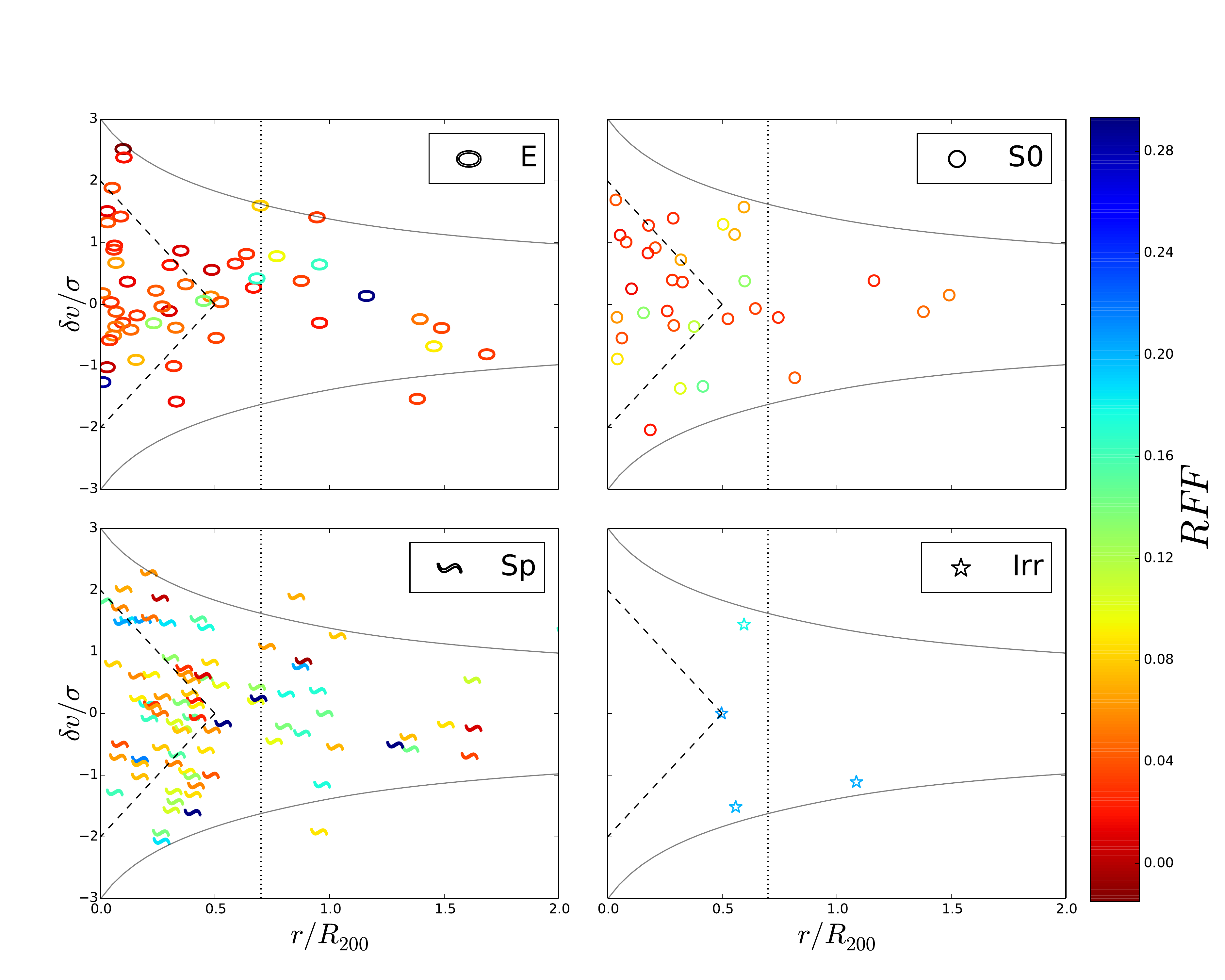}
 \caption[$RFF$ distribution on the PPS for galaxies of each morphological type]{\label{pps_mrff} $RFF$ distribution on the PPS for galaxies of each morphological type. Similar to Figures~\ref{pps_rff} \& ~\ref{pps_ares}, the dashed boundary encompasses the `virial' region. Although the majority of early-type galaxies (E+S0) are smooth, they show equal fractions of rough and smooth galaxies in the virial and infall regions of the cluster. For the spiral galaxies, the majority are rough with no obvious dependence on the local environment.}   
\end{figure*}

\begin{figure*}
 
 \hspace{-1.2cm}
 \includegraphics[width=1\textwidth]{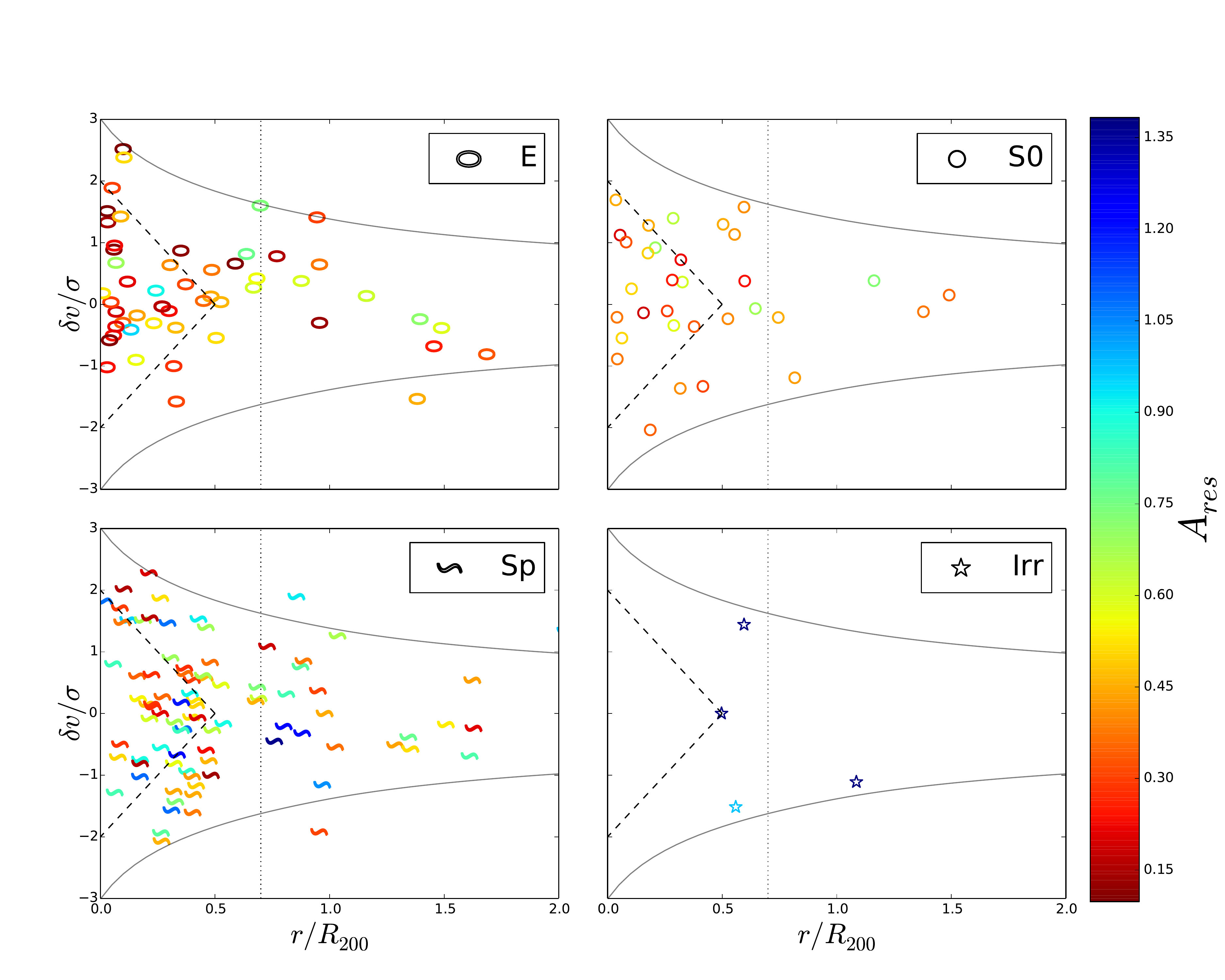}
 \caption[$A_{\rm res}$ distribution on the PPS for galaxies of each morphological type]{\label{pps_mares}Similar to Figure~\ref{pps_mrff}, $A_{\rm res}$ distribution on the PPS for galaxies of each morphological type. Although the majority of early-type galaxies (E+S0) are symmetric, they show equal fractions of asymmetric and symmetric galaxies in the virial, and infall region of the cluster. For the spiral galaxies, the majority are asymmetric with no obvious dependence on the local environment.}   
\end{figure*}

The PPS thus constructed provides a unique snapshot into the clusters assembling their galaxies at $\langle{z}\rangle\sim0.65$. From Figures~\ref{pps_rff} and ~\ref{pps_ares}, the variation in morphology across the PPS is prominent, and masks out any possible variation which may exist for $RFF$ and $A_{\rm res}$. Figures~\ref{pps_mrff} and ~\ref{pps_mares} show the PPS for each morphological type separately, still colour-coded by $RFF$ and $A_{\rm res}$. The fractions tabulated in Table~\ref{table2} confirm that the early-type galaxies are spread across the cluster virial and infall regions, while the late-type galaxies are preferentially located outside the virial region. 

\subsection[$RFF$ and $A_{\rm res}$ as a function of location on the PPS]{$RFF$ and $A_{\rm res}$ as a function of location on the PPS}  
\label{ssec:pps-sfh-morph}
In order to study the implications of galaxy structure on such a cluster-wide scale, we next look into the distribution of $RFF$ and $A_{\rm res}$ on the PPS for each morphology type. To quantify the distribution of $RFF$ on the PPS, we divide the galaxy sample into `rough'/`smooth' galaxies based on the boundary in $RFF$ distribution encompassing majority of the early-type (E+S0) galaxies (K17). This boundary was qualitatively determined using the median $RFF$ of E+S0 galaxies as an initial approximation. Following this, galaxies with $RFF > 0.05$ formed the subclass of `rough' galaxies while galaxies with $RFF \leqslant 0.05$ formed the subclass of `smooth' galaxies. Using a similar approach to the $A_{\rm res}$ distribution, we classified galaxies as `asymmetric' if they displayed $A_{\rm res} > 0.5$ while galaxies with $A_{\rm res} \leqslant 0.5$ were considered `symmetric'. Table~\ref{table3} gives the fractions of quantitative structure for each morphology type in the virial and infalling regions as well for the field environment.      

\begin{table*}
  \begin{center}
  
   \centering
   \caption[The relative fractions for galaxies defined as rough, smooth, symmetric and asymmetric,
    for a fixed morphology in different cluster regions and the field]{\label{table3} The relative fractions for galaxies defined as rough, smooth, symmetric and asymmetric as described in Section~\ref{ssec:pps-sfh-morph},
    for a fixed morphology in different cluster regions and the field. The errors on fractions denote the $1\sigma$ uncertainty computed over Wilson interval (cf. Section~\ref{ssec:pps-construct}).  \newline}
    \begin{tabular}{llccc}

    \centering
    \textbf{Morphology} &  & \textbf{Virial region} & \textbf{Infall region} & \textbf{Field} \\
  
     & & $r \lesssim 0.5R_{\rm 200}$ & $r > 0.5R_{\rm 200}$ &   \\
     & & $|\delta v| \lesssim 2\sigma_{\rm cl}$& $|\delta v| > 2\sigma_{\rm cl}$ & \\
    \hline
    \hline
    \multicolumn{5}{c}{}\\
    \multirow{7}{20pt}{Ellipticals (E)} & Smooth & 0.71$\pm$0.08 & 0.69$\pm$0.09 & 0.69$\pm$0.11  \\ 
    & Rough & 0.29$\pm$0.08 & 0.31$\pm$ 0.09 & 0.31$\pm$0.11  \\ 
    \multicolumn{5}{c}{}\\
    & Symmetric & 0.74$\pm$0.08 & 0.61$\pm$0.09 & 0.64$\pm$0.11 \\ 
    & Asymmetric & 0.26$\pm$0.08 & 0.39$\pm$0.09 & 0.36$\pm$0.11 \\ 
    \multicolumn{5}{c}{}\\
    & Young &0.05$\pm$0.04 & 0.02$\pm$0.02& 0.14$\pm$0.08\\ 
    & Intermediate & 0.09$\pm$0.05 & 0.28$\pm$0.09 & 0.31$\pm$0.11 \\ 
    & Old & 0.88$\pm$0.06 & 0.72$\pm$0.09 & 0.58$\pm$0.12 \\ 
  
    \hline
    \multicolumn{5}{c}{}\\
    \multirow{7}{20pt}{Lenticulars (S0)} & Smooth & 0.74$\pm$0.11 &0.50$\pm$0.12  & 0.64$\pm$0.18  \\ 
    & Rough & 0.26$\pm$0.11 & 0.50$\pm$0.12 & 0.36$\pm$0.18 \\
    \multicolumn{5}{c}{}\\
    & Symmetric & 0.68$\pm$0.11& 0.79$\pm$0.10& 0.79$\pm$0.15\\
    & Asymmetric & 0.32$\pm$0.11&0.21$\pm$0.10 & 0.21$\pm$0.15\\
    \multicolumn{5}{c}{}\\
    & Young &0.03$\pm$0.03 & 0.03$\pm$0.03& 0.07$\pm$0.07\\
    & Intermediate &0.21$\pm$0.10 & 0.26$\pm$0.11&0.07$\pm$0.07 \\
    & Old &0.79$\pm$0.10 &0.74$\pm$0.11 & 0.93$\pm$0.07\\
    
    \hline 
    \multicolumn{5}{c}{}\\
    \multirow{7}{20pt}{Spirals (Sp)} & Smooth &0.16$\pm$0.07 &0.16$\pm$0.05 & 0.12$\pm$0.05\\
    & Rough &0.84$\pm$0.07 &0.84$\pm$0.05 &0.88$\pm$0.05 \\
    \multicolumn{5}{c}{}\\
    & Symmetric & 0.40$\pm$0.09& 0.47$\pm$0.06& 0.37$\pm$0.07 \\
    & Asymmetric & 0.60$\pm$0.09& 0.53$\pm$0.06&0.63$\pm$0.07 \\
    \multicolumn{5}{c}{}\\
    & Young & 0.19$\pm$0.07& 0.37$\pm$0.06& 0.52$\pm$0.07\\
    & Intermediate &0.33$\pm$0.09 &0.40$\pm$0.06 & 0.35$\pm$0.07\\
    & Old &0.50$\pm$0.09 &0.24$\pm$0.05 & 0.14$\pm$0.05\\
    \hline
    \end{tabular}
  \end{center}
\end{table*}

The elliptical galaxies show equal fractions of smooth, rough, symmetric and asymmetric galaxies across all cluster regions, and the field. This trend is also mirrored in the fractions observed for the lenticular galaxies. The fact that lenticular galaxies show no significant changes in structure irrespective of their position on the PPS relates to the observation by \citet{jaffe11b} that cluster ellipticals and lenticulars are found to have similar stellar population ages inferred from their colours. We would thus expect them to be equally smooth and symmetric. As evident from the lower left panels in Figure~\ref{pps_mrff} and ~\ref{pps_mares} for spiral galaxies and Table~\ref{table3}, most spirals in the cluster are rough and asymmetric. However, there is no obvious difference in the $RFF$ and $A_{\rm res}$ distributions of spiral galaxies across environments.   

We also look at the variation of visual disturbance classes (K17; Table~\ref{table3}) for spirals in various cluster regions and the field. The majority of the spirals in the virial region are found to be visually undisturbed as compared to the cluster outskirts and the field.  In contrast, the fraction of externally disturbed spiral galaxies does not seem to depend on the local environment. This observation supports the lack of environmental dependence of visual disturbance classes, as found in K17. In particular, we note in K17 that $A_{\rm res}$ is sensitive to mergers which are relatively harder to occur in dense cluster environments. Although not a part of this paper, this premise has been extensively explored in the recent EDisCS paper by \citet{deger18}, where they combine the visual disturbance classifications from K17 with alternative quantitative measures of Gini (G) and M20, which together are efficient in merger detection. Exploring the global and local environment (through PPS analysis), Deger et al. report that the fraction of tidal interactions and mergers ($f_{TIM}$) shows no significant trend as a function of environment, except for a mild boost at intermediate/group environments and a probable enhancement at very high densities. A probable explanation for the lack of any significant trends is that a slower process like morphological transformation would occur on timescales greater than cluster crossing time ($\sim2$ Gyrs, see Section \ref{ssec:pps-sfh-ages}) and hence the signatures would get washed out in the PPS. A similar scenario was proposed by \citet{jaffe11}, who, using the same galaxy sample, found that, while kinematical disturbances in the gas present in galaxies are correlated with the environment, the stellar light distribution remains independent of environment, suggesting longer timescales for structural transformation as compared to variation in stellar ages.

\begin{figure*}
 
 \hspace{-1.5cm}
 \includegraphics[width=1.1\textwidth]{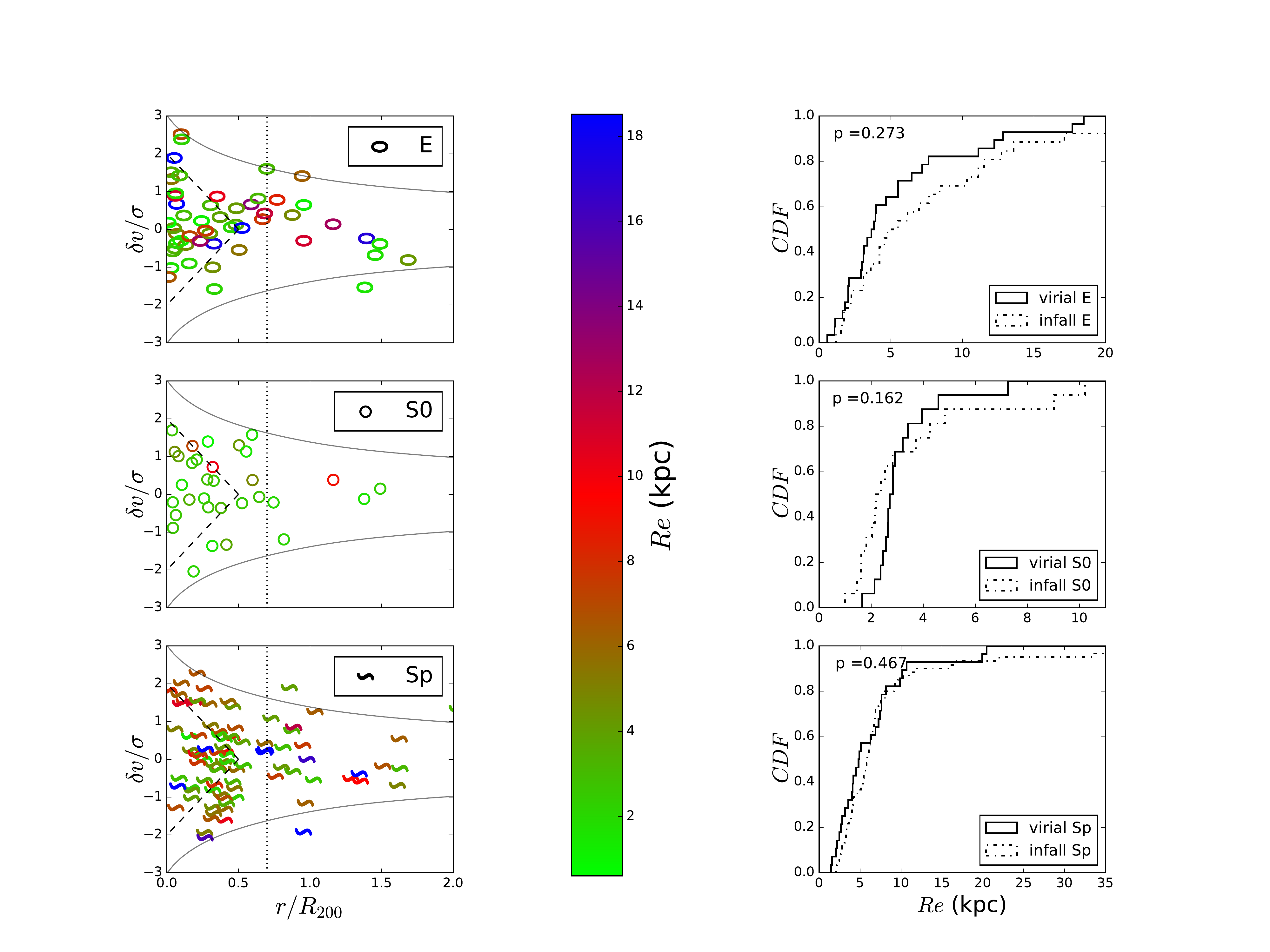}
 \caption[$Re$ distribution on the PPS for galaxies of each morphological type]{\label{pps_re} $Re$ distribution on the PPS for galaxies of each morphological type. The panels on the right denote the cumulative distribution functions of $Re$ for fixed morphologies, with the solid lines indicating virial galaxies and the dashed lines represnting the infall galaxies. The inset p-values are significance results from the two-sample K--S tests performed under the null hypothesis that the virial and infall galaxies are drawn from the same $Re$ distribution. Although the range of $Re$ are different for galaxies of each morphology type, their sizes show no obvious dependence on the local environment.}   
\end{figure*}

We briefly revisit the global structural properties of galaxies like single S\'{e}rsic indices, galaxy sizes and bulge-to-total luminosity ratio (B/T), and investigate their correlation with the position on PPS. \citet{kelkar15} demonstrate that the sizes of EDisCS galaxies appear to not depend on the global cluster-field environment, irrespective of being compared over fixed morphology or galaxy colours. Building on the sample utilised in \citet{kelkar15}, we present a comparison of galaxy sizes ($Re$) over the internal cluster environment ie. within and outside virial radius as defined in Section~\ref{sec:pps}, in Figure~\ref{pps_re}. We also demonstrate results of two-sample K--S tests perfomed to check the significance of the variation in $Re$ across the virial and infall regions, under the null hypothesis that each sample of these cluster galaxies are drawn from the same $Re$ distribution. Complementing Figure 2 from \citet{kelkar15}, our analysis further establishes that galaxies indeed have similar sizes across the varied cluster environment, at a fixed morphology.\

\begin{figure*}
 
 \hspace{-1.5cm}
 \includegraphics[width=1.1\textwidth]{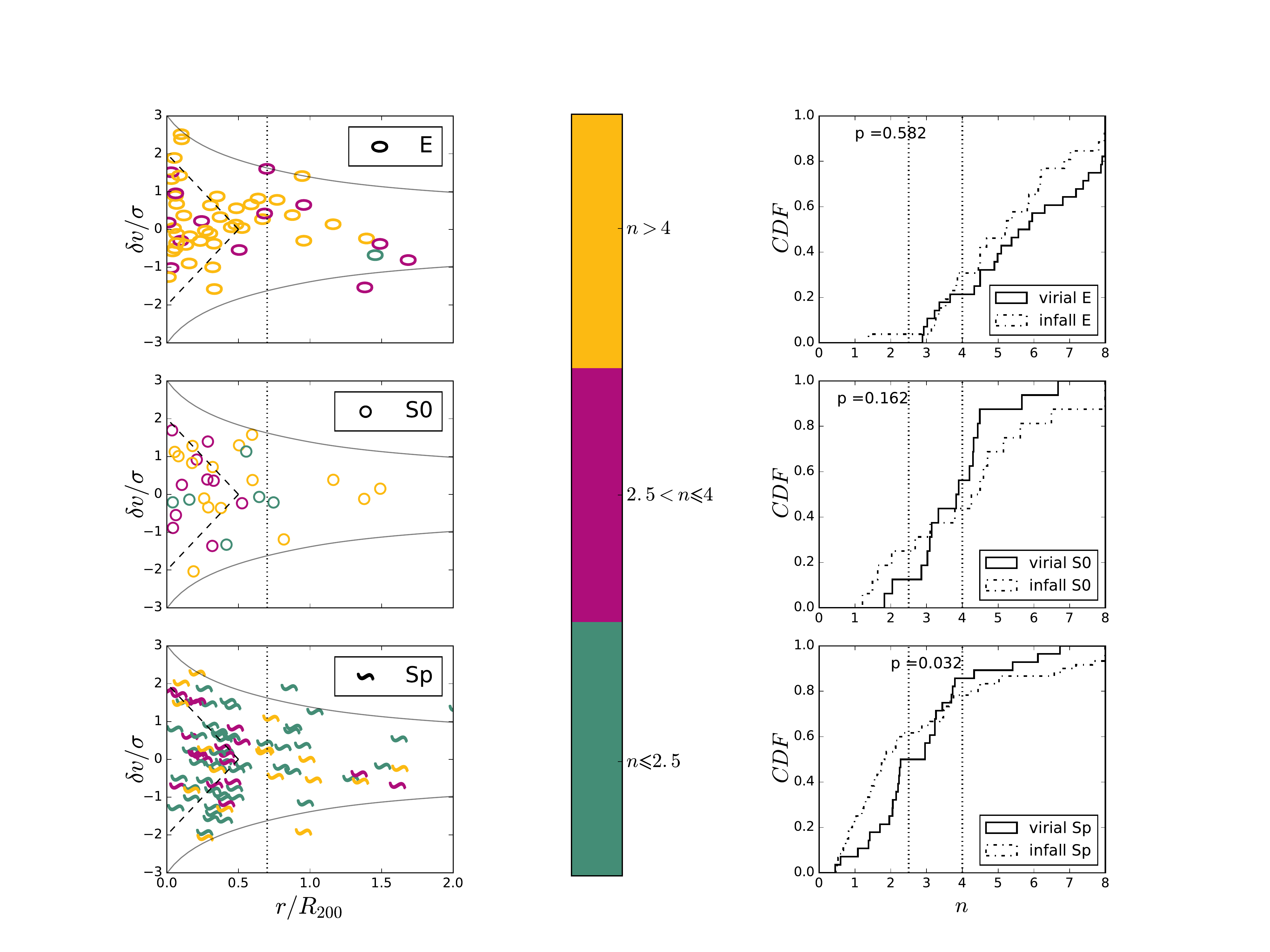}
 \caption[The distribution of S\'{e}rsic index $n$ on the PPS for galaxies of each morphological type]{\label{pps_n} The distribution of S\'{e}rsic index $n$ on the PPS for galaxies of each morphological type. The panels on the right denote the cumulative distribution functions of $n$ for fixed morphologies, with the solid lines indicating virial galaxies and the dashed lines representing the infall galaxies. The vertical dotted lines indicate the universal categories into which the $n$ is shown on the PPS. The inset p-values are significance results from the two-sample K--S tests performed under the null hypothesis that the virial and infall galaxies are drawn from the same $n$ distribution. Majority of ellipticals show $n>4$ while most of the spirals have $n\leqslant2.5$. Interestingly, the infalling cluster spirals show smaller $n$ as compared to spiral galaxies in the central regions of the stacked cluster suggesting a possible dependence of $n$ on internal cluster environment.}   
\end{figure*}

\begin{figure}
 
 \hspace{-1.5cm}
 \includegraphics[width=1\textwidth]{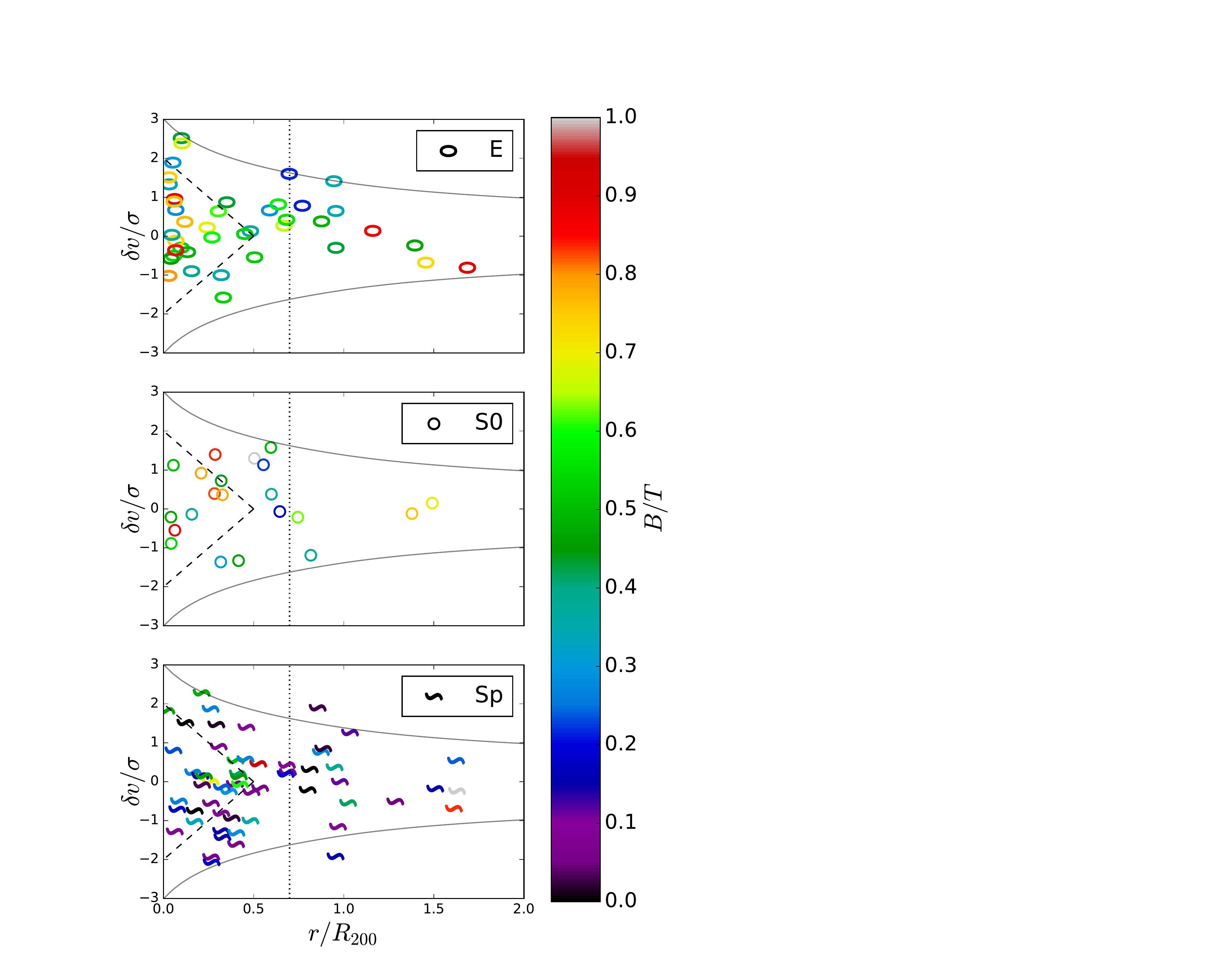}
 \caption[The distribution of bulge-to-total luminosity ratio ($B/T$) on the PPS for galaxies of each morphological type]{\label{pps_bt} The distribution of bulge-to-total luminosity ratio ($B/T$) on the PPS for galaxies of each morphological type. While the ellipticals are shown to have highest B/T, the overall variation of B/T across the PPS, for a fixed morphology, is insignificant. This indicates that the effects of internal cluster environment, if any, do not strongly affect the bulge of galaxies irrepective of their moprphologies.}   
\end{figure}

The systematic variation in S\'{e}rsic index ($n$) of galaxies, at a fixed stellar mass, could robustly hint at the merging and strong gravitational interactions which the galaxies underwent. We thus split our sample into universal categories based on S\'{e}rsic indices, namely `disks' ($n\leqslant2.5$), and `spheroids' ($2.5<n\leqslant4$). Note however, that due to the asymptotic nature of the single S\'{e}rsic profile for large $n$, a third category of galaxies is defined with $n\geqslant4$. Figure ~\ref{pps_n} shows the PPS of galaxies for each morphology type, colour-coded according to these broad structural categories. Similar to $Re$ distribution on PPS, we employ the two sample K--S test to statistically quantify the variation in the distribution of `disk' and `spheroid' galaxies across the virial, and infall regions of PPS. While the mix of disky/spheroidal galaxies is different for different morphology types, the majority of cluster ellipticals have $n>4$ which is unsurprising as cluster ellipticals, especially in the core, are brighter and massive, and hence are expected to have higher $n$. Moreover, the $n$ distribution for the ellipticals in the virial and infall regions is similar. This result is somewhat contradictory to the observation made by \citet{d'onofrio15} who demonstrate that the cluster ellipticals at EDisCS redshifts, at a fixed stellar mass, consistently show lower $n$ at smaller cluster-centric radius suggestive of loss of stellar envelope in galaxies at such high densities. Note however that our analysis uses a 2-D approach applied over a limited mass range which could be a reason for the apparent lack of trend. The spirals in the virial region, on the other hand, marginally have higher $n$ while spirals in infall regions tend to show low $n$, with a confidence $>95\%$ (lower right panel of Figure~\ref{pps_n}). This suggests that the spirals in the densest cluster environment are more bulge-dominated as compared to rest of the disky cluster spirals. This result has important connotations on the reported structural evolution of cluster spirals to bulge-dominated lenticular galaxies via plausible environmental channels. To explore this further, we look into the bulge-to-total luminosity ratio (B/T) \citep[\textsc{gim2d};][]{simard10, saglia10} of our galaxies as a function of PPS (Figure~\ref{pps_bt}). Surprisingly, we report that the B/T distribution in the virial and infall regions is similar irrespective of their morphology, with the ellipticals and lenticulars having higher B/T in general than the cluster spirals. The modest increase in the $n$ of the cluster spirals in virial region and the subsequently unchanged B/T across the internal cluster environment, may hint towards a physical processes like tidal stripping affecting the outermost stars in the disk of infalling spirals thus lowering their $n$ as opposed to bulge-building merger processes in core spirals. Alternatively, the core galaxies morphologically classified as spirals may not have grown their bulges enough to be detected by our measurement.

In summary, galaxies of the same morphology in the virial region are not different from the infalling galaxies with respect to their $RFF$ and $A_{\rm res}$, though there is a hint of environmental trend in the distribution of $n$. These observations suggest there are no significant changes in the structure of galaxies irrespective of their position on the PPS; at the very least these changes perhaps are not observable on short timescales.

\begin{table}
  \begin{center}
  
   \centering
   \caption[The relative fractions for spiral galaxies identified in each disturbance class, in different cluster regions and the field]{\label{table4} The relative fractions for spiral galaxies identified in each disturbance class (cf. K17), in different cluster regions and the field. As described in Table~\ref{table3}, the errors on fractions denote the $1\sigma$ uncertainty (cf. Section~\ref{ssec:pps-construct}). \newline}
    \begin{tabular}{lccc}

    \centering
    \textbf{Spirals}  & \textbf{Virial region} & \textbf{Infall region} & \textbf{Field} \\
  
      & $r \lesssim 0.5R_{\rm 200}$ & $r > 0.5R_{\rm 200}$ &   \\
      & $|\delta v| \lesssim 2\sigma_{\rm cl}$& $|\delta v| > 2\sigma_{\rm cl}$ & \\
    
    \hline
    \hline
    \multicolumn{4}{c}{}\\
    Undisturbed (0) &0.57$\pm$0.09 & 0.40$\pm$0.06& 0.41$\pm$0.07\\
    Internally Asymmetric (iA)  & 0.19$\pm$0.07&0.19$\pm$0.05 & 0.20$\pm$0.06\\
    Interacting (I) &0.12$\pm$0.06 &0.19$\pm$0.05 &0.20$\pm$0.06 \\
    Tidal (T) &0.05$\pm$0.04 &0.12$\pm$0.04 &0.16$\pm$0.05\\
    Mergers (M) &0.09$\pm$0.05 & 0.11$\pm$0.04& 0.05$\pm$0.03\\
    \hline
    \end{tabular}
  \end{center}
\end{table}

\subsection[Stellar ages as a function of location on the PPS: a glimpse at accumulating passive galaxies]{Stellar ages a function of location on the PPS: a glimpse at accumulating passive galaxies}  
\label{ssec:pps-sfh-ages}

\begin{figure*}
 
 \hspace{-1.2cm}
 \includegraphics[width=1\textwidth]{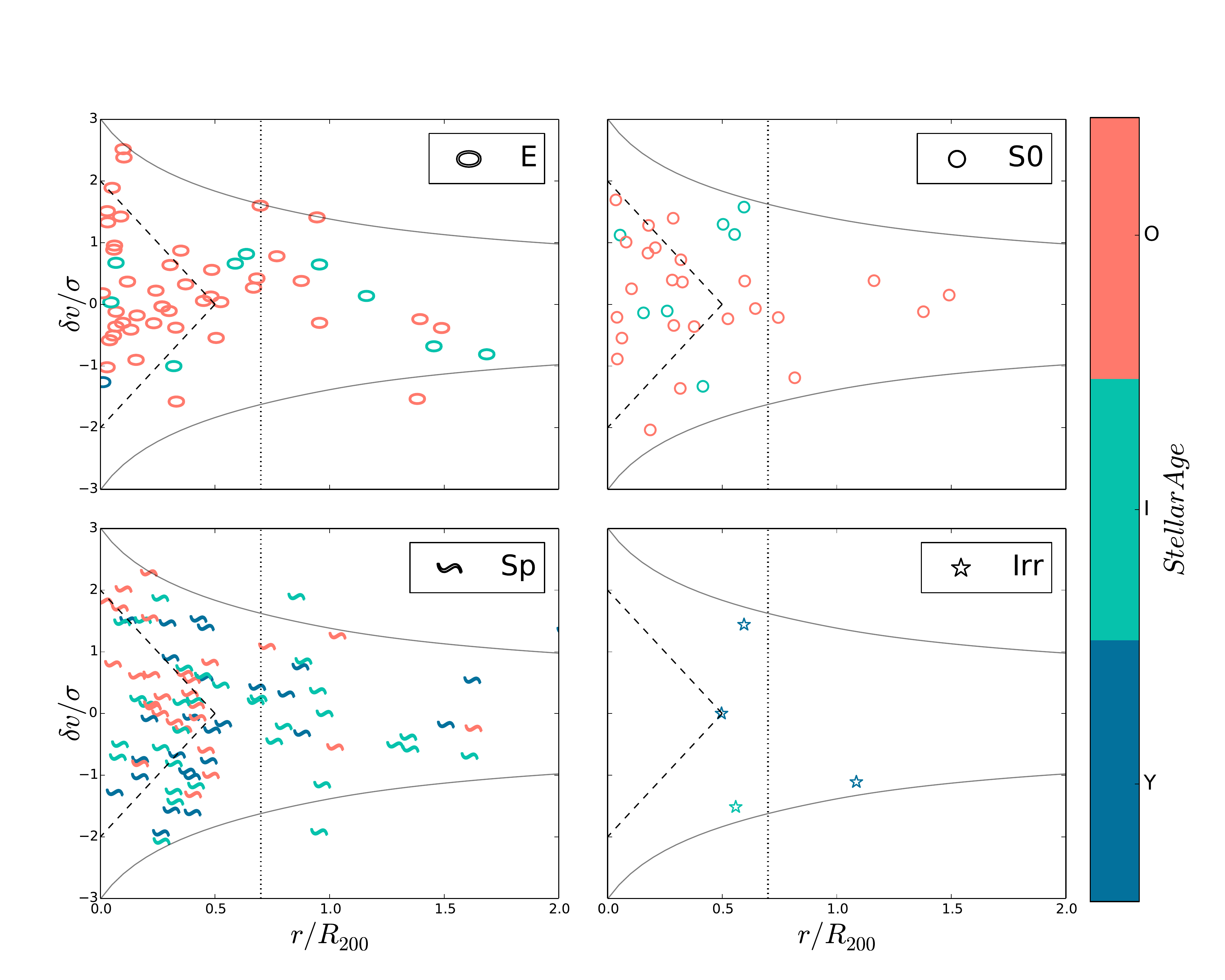}
 \caption[Distribution of stellar ages on the PPS for galaxies of each morphological type]{\label{pps_age} Distribution of stellar ages on the PPS for galaxies of each morphological type. As expected, `old' early-type galaxies populate the virial region while the `intermediate' galaxies seem to avoid the cluster core. This trend is enhanced for cluster spirals where `old' spirals are found in the cluster core, and spirals with younger stellar population are found in the cluster outskirts.}   
\end{figure*}

We next investigate the ages of stellar populations in galaxies as a function of the location on the PPS. Figure~\ref{pps_age} shows the PPS for the galaxies of each morphological type colour-coded according to the age of their stellar populations (cf. Section~\ref{sec2:sfh}).  In contrast to the lack of structural trends with environment, several conclusions can be drawn regarding trends with age.  

The majority of the elliptical galaxies are `old' and tend to populate the virialized regions while the `intermediate' age ellipticals avoid the cluster core. This observation is further supported by the fractions of ellipticals with old, intermediate, and younger stellar ages in the cluster regions and the field (cf. Table~\ref{table3}). Note however that some of the old elliptical galaxies seen in the infall region may have been pre-processed or internally quenched, while some will be older cluster members that are not observed in the virialised triangle due to scatter in the PPS and contamination \citep{rhee17}.

The lenticular galaxies, however, show the interesting property that the vast majority of the field lenticulars harbour older stellar populations. In all cluster environments (the virial and infall regions), `old' lenticulars are seen to be dominant but there are some cluster lenticulars with intermediate stellar ages. Although the number of field lenticulars is small, this could suggest that the evolutionary history of field and cluster lenticulars may be different. The transformation of spirals into lenticulars is being environmentally driven, while the field lenticulars could originate from spirals whose gas supply has been exhausted or through minor mergers \citep{arnold11}.

For the spiral galaxies, the fraction of `old' spirals increases towards the core, indicating the build up of older populations in clusters. This is clearly a cluster phenomenon originating, perhaps, on the cessation of star formation as the galaxies enter cluster core, where they remain and become old.  
On the other hand, the fraction of spirals with younger stellar populations shows a complementary increase towards the field. This observation likely supports the findings from recent works like \citet{biviano16}, where they observe that passive and star-forming galaxies have similar elongated orbits with increasing clustercentric radii at intermediate redshift. Moreover, these orbits are similar to those of star-forming galaxies in $z=0$ clusters, suggesting a prominent orbital evolution for the passive galaxies with time. Interestingly, the fraction of `intermediate' age spirals remains approximately constant over all environments. While a fraction of these `intermediate' age spirals may truly denote the spiral galaxies in transition at these redshifts, one must be careful while considering these spirals as their incidence in different cluster regions may partially be attributed to projection effects.

The presence of `intermediate' age cluster spirals and lenticulars, and the higher fraction of `old' spirals in the cluster core illustrate the proliferation of passive old galaxies in cluster cores. Consistent fractions of `intermediate' age spirals across the infall and core regions would arise if spiral galaxies are able to form stars for a significant amount of time after infalling into the cluster. Considering our stacked cluster from the PPS analysis, it takes $\sim2$ Gyrs for a galaxy with a peculiar velocity $\sim600$ km/s to reach the cluster core from the virial radius, thereby suggesting an environmental process acting over this timescale. The possibility of such prolonged mechanism affecting star formation in cluster spirals thus rules out rapid shutdown of star formation in spirals upon infalling into clusters \citep[See ][]{cantale16}. With the observed lack of strong trends in the $RFF$ and $A_{\rm res}$ distributions across the PPS, galaxy starvation stands as likely candidate which can deplete the hot gas in these cluster galaxies for a longer period of time without affecting intrinsic galaxy structure. 
 
The build up of passive `old' spirals in cluster cores, however, suggests two possibilities: that either galaxies encountering the dense ICM in cluster cores most likely experience rapid removal of cold gas through mechanisms like Ram pressure stripping, eventually turning them passive, or that they have been in the cluster for a long time. In both cases, the global galaxy structure remains intact as substantiated by the insignificant dependence of $RFF$ and $A_{\rm res}$ as a function of position on the PPS. This scenario is supported by complementary results by \citet{rudnick17} who demonstrate that older cluster galaxies with weak [O\textsc{ii}] emission avoid the core region in clusters. The end result in either case would thus be observed as the accumulation of `old' spirals in cluster cores. The associated morphological transition of `old' spirals into early-type galaxies happens gradually with the `old' spirals slowly losing the star formation in their disks due to loss of gas, and possibly transforming into lenticulars. These lenticulars may show a final episode of star formation in their centres \citep{johnston14} building a larger bulge.
 
\section[Conclusions]{Conclusions}  
\label{sec:conc}

In this paper, we present a study of the putative links that may exist between the internal properties of galaxies, such as their morphology, structure and star formation history (SFH), and the different cluster environments identified using the projected phase--space (PPS) for a sample of intermediate redshift ($0.4<z<0.8$) cluster and field galaxies from the ESO Distant Cluster Survey (EDisCS). 
 
Using HST $I-$band images (rest-frame $\sim B$) we fit simple smooth symmetric models to the galaxies' surface brightness profiles and measure deviations of the galaxies' light distributions from these models with two parameters, $A_{\rm res}$ (asymmetry of the residuals from the model fits, `asymmetry' for short) and $RFF$ (residual flux fraction, or the fraction of the total galaxy light present in the residuals, measuring `roughness').
 
Using VLT multi-object spectroscopy, we derive age-sensitive spectral indicators ($H_{\delta\rm{A}}$, $H_{\gamma\rm{A}}$ and $D_{n}4000$) to obtain information on the star-formation histories of the galaxies. High-quality spectroscopic redshifts are used to build projected phase--space diagrams for a subset of the clusters. These are plots of the radial velocities of the cluster galaxies in the rest-frame of the cluster normalised by its velocity dispersion ($\sigma_{\rm clus}$) vs.\ their projected radial distances from the cluster centre in units of its virial radius ($R_{200}$). To improve the statistics, we stack these diagrams and build an average PPS diagram for these clusters.
 
Analysing the galaxies' properties in conjunction with their environmental information derived from the PPS we reach the following conclusions:
    
\begin{enumerate}
  \item The quantitative structural measurements $A_{\rm res}$ and $RFF$  correlate very well with the age of the galaxies' stellar populations derived from the ($H_{\delta\rm{A}} + H_{\gamma\rm{A}})/2$ vs.\ $D_{n}4000$ plane. At a fixed morphology, young, star-forming galaxies are consistently rougher and more asymmetric than older ones. $A_{\rm res}$ and $RFF$ become progressive lower with time since the latest episodes of star formation.
  \item We find a significant correlation between the position of the galaxies on the PPS and their stellar ages, irrespective of their morphology. We also observe an increasing fraction of galaxies with older stellar populations towards the cluster core, but parallel trends of the galaxies' structural parameters ($A_{\rm res}$ and $RFF$) with PPS location are not observed.
  \item We report that infalling spirals marginally have low single S\'{e}rsic index as compared to spiral galaxies in cluster core.   
  \item Incorporating the visual information provided by \cite{kk17} on the degree of disturbance observed in the galaxies' HST images, we find that the virial region of the cluster contains a higher fraction of visually undisturbed spirals than the cluster outskirts and the field.
\end{enumerate}
 
These results have important implications for the overall picture of galaxy transformation within the context of environmental effects. The dwindling fraction of spirals with young stellar populations in the cluster core suggests that, as a galaxy is accreted into the virialised region, its star-formation is reduced and its stellar population ages. Since galaxy morphology and quantitative measurements of galaxy structure such as $RFF$ and $A_{\rm res}$ correlate with SFH, we would naively expect that corresponding trends between these parameters and location in PPS should also  be observed. The fact that these trends are not observed suggests that the timescales involved in the suppression of star formation of cluster galaxies are significantly shorter that the timsecales for their structural and morphological transformation, in agreement with the findings of \cite{kk17}. In other words, galaxies start becoming structurally smoother and more symmetric as their star formation is reduced and ceases, but their morphological transformation happens later \citep[see also][]{wolf09,jaffe11}.
 
Given the limitations of the sample and available data, and the lack of information on the properties and distribution of the intra-cluster medium, it is not possible to know with any certainty what the actual physical causes of the proposed transformations are. However, we can confidently say that galaxies with old stellar populations are structurally smoother and more symmetric, but the actual morphological change will follow later, once the star formation has already been truncated. The most obvious effect of the environment is the accumulation of the old early-type passive galaxies in the cluster core. We speculate that the explanation for this could either be the result of relatively gentle cluster-specific mechanisms such as ram-pressure stripping and galaxy starvation, or the result of the cluster buildup history. Or, more likely, a combination of both.

\section*{Acknowledgments}

Based on observations made 
with the NAS/ESA \textit{Hubble Space Telescope}, obtained at the Space Telescope Science Institute, which is operated by the Association of Universities for Research in 
Astronomy, Inc., under NASA contract NAS 5-26555. These observations are associated with proposal 9476. Support for this proposal was provided by NASA through grant
from the Space Telescope Science Institute.

Based on observations obtained at the ESO Very Large Telescope (VLT) as a part of the Large Programme 
166.A-0162

We acknowledge the EDisCS team for providing useful comments on the this paper. We would also like to thank the Referee for the valuable feedback which contributed positively to the work presented in this paper. GHR acknowledges the support of NASA grant HST-AR-12152.001-A, NSF grants 1211358 and 1517815, the support of an ESO visiting fellowship, and the hospitality of the Max Planck Institute for Astronomy, the Max Planck Institue for Extraterrestrial Physics, and the Hamburg Observatory. GHR also acknowledges the support of a Alexander von Humboldt Foundation Fellowship for experienced researchers. GHR recognizes the support of the International Space Sciences Institute for their workshop support.Y.J. acknowledges financial support from CONICYT PAI (Concurso Nacional de Inserci\'on en la Academia 2017) No. 79170132 and FONDECYT Iniciaci\'on 2018 No. 11180558.



\appendix

\bsp

\label{lastpage}

\end{document}